\documentclass[5p]{elsarticle}
\usepackage{amsfonts}
\usepackage{amsmath}
\usepackage{multirow}
\usepackage{amssymb}
\usepackage{textcomp}
\usepackage{algorithm}
\usepackage{algpseudocode}
\usepackage{subfig}
\usepackage{graphicx}
\usepackage{xcolor}
\usepackage{multirow}
\usepackage{booktabs}
\biboptions{authoryear}
\journal{arXiv.org}

\DeclareMathOperator*{\argmin}{arg\,min}
\newcommand\norm[1]{\left\lVert#1\right\rVert}

\def\NoNumber#1{{\def\alglinenumber##1{}\State #1}\addtocounter{ALG@line}{-1}}
\algnewcommand\Or{\textbf{or} }

\begin{document}
\sloppy
\begin{frontmatter}

\title{Multi-compartment diffusion-relaxation MR signal representation in the spherical 3D-SHORE basis}

\author[AGH]{Fabian Bogusz}
\ead{fbogusz@agh.edu.pl}
\address[AGH]{AGH University of Kraków, Kraków, Poland}

\author[UVa]{Tomasz Pieciak\corref{cor1}}
\ead{tpieciak@tel.uva.es}
\address[UVa]{Laboratorio de Procesado de Imagen (LPI), ETSI Telecomunicaci\'{o}n, Universidad de Valladolid, Valladolid, Spain}

\cortext[cor1]{Corresponding author}

\begin{abstract}
Modelling the diffusion-relaxation magnetic resonance (MR) signal obtained from multi-parametric sequences has recently gained immense interest in the community due to new techniques significantly reducing data acquisition time. A preferred approach for examining the diffusion-relaxation MR data is to follow the continuum modelling principle that employs kernels to represent the tissue features, such as the relaxations or diffusion properties. However, constructing reasonable dictionaries with predefined signal components depends on the sampling density of model parameter space, thus leading to a geometrical increase in the number of atoms per extra tissue parameter considered in the model. That makes estimating the contributions from each atom in the signal challenging, especially considering diffusion features beyond the mono-exponential decay.

This paper presents a new Multi-Compartment diffusion-relaxation MR signal representation based on the Simple Harmonic Oscillator-based Reconstruction and Estimation (MC-SHORE) representation, compatible with scattered acquisitions. The proposed technique imposes sparsity constraint on the solution \textit{via} the $\ell_1$ norm and enables the estimation of the microstructural measures, such as the return-to-the-origin probability, and the orientation distribution function, depending on the compartments considered in a single voxel. The procedure has been verified with in silico and in vivo data and enabled the approximation of the diffusion-relaxation MR signal more accurately than single-compartment non-Gaussian representations and multi-compartment mono-exponential decay techniques, maintaining a low number of atoms in the dictionary. Ultimately, the MC-SHORE procedure allows for separating intra-/extra-axonal and free water contributions from the signal, thus reducing the partial volume effect observable in the boundaries of the tissues.
\end{abstract}

 
\end{frontmatter}

%----------------------------------------------------------------------------------------------
%----------------------------------------------------------------------------------------------
%----------------------------------------------------------------------------------------------
\section{Introduction}
\label{sec:introduction}
The multi-parametric sequences facilitate acquiring the magnetic resonance (MR) signal representing complementary biophysical and chemical processes. An illustrative example of multi-contrast acquisition is a combined diffusion-relaxometry MR imaging that disentangles the medium's diffusion and relaxation properties \citep{benjamini2016use,kim2017diffusion,hutter2018integrated,ning2019joint,nagtegaal2020fast,slator2021combined}. While the diffusion process reflects the presence of biological barriers that restrict and hinder the movements of water molecules, the relaxation illustrates the chemical composition of the imaged tissue \citep{does2018inferring,afzali2021sensitivity}. Recent advances in diffusion-relaxation MR sequences have boosted the collection of sparsely sampled MR data, moving this imaging technique towards clinically acceptable time regimes \citep{hutter2018integrated,fritz2021mesmerised,leppert2021efficient}.

To retrieve quantifiable and valuable parameters from a diffusion-relaxation MR acquisition, one can employ diverse approaches to characterize the multi-contrast signal, such as the biophysical modelling, signal representation or continuum modelling \citep{slator2021combined}. The biophysical models assume the MR signal comprises a finite and preferably low number of complementary compartments that reflect the underlying tissue microstructure \citep{jelescu2020challenges}. The signal representation expresses the MR signal using a cumulant expansion or decomposes it using a predefined set of basis functions \citep{ozarslan2009simple,hosseinbor2013bessel,ning2015estimating,zucchelli2016lies,wang2024q}. The diffusion tensor imaging (DTI) \citep{basser1994mr} and diffusion kurtosis imaging \citep{jensen2005diffusional} are classic examples of signal representation methods used in a clinical scenario nowadays. The last, the continuum modelling assumes the MR signal can be represented using the Fredholm equation of the first kind \citep{slator2021combined}, and therefore characterised using the continuous kernel function that is a simple model for $T_1$ or $T_2/T_2^*$ relaxations, or the diffusion evolution.

The prevalence of techniques based on continuum modelling assumes a mono-exponential kernel to model the diffusion process \citep{benjamini2016use,kim2017diffusion}. Such an approach is reasonable under the assumption of isotropy of the diffusion process. However, once moving towards a higher b-value regime, this assumption starts to be violated. The kurtosis expansion \citep{jensen2005diffusional} can reduce the error of the signal approximation but at the cost of including a new parameter that needs to be sampled for dictionary construction. Besides, the higher the sampling of the kernel parameter space, the more atoms the dictionary contains, which also leads to an increased numerical complexity of the inverse problem.

Multi-parametric acquisitions can be time-consuming as the MR signal is measured using different combinations of acquisition protocol parameters. To characterise the MR signal from a reduced number of acquisitions, \cite{benjamini2016use} proposed a regularization that uses the projections of the 2D diffusion-relaxation spectrum onto the 1D separate diffusion or relaxation spectra. The estimation of the 2D spectrum promotes the solutions that are compatible with the respective 1D spectra. \cite{kim2017diffusion} has introduced the spatial smoothness constraint that promotes the similarity of the correlation spectra estimated for neighbouring voxels. Another approach by \cite{nagtegaal2020fast} engages the squared $\ell_1$ norm that allows rewriting the standard two-term objective into a one-term function.
The method uses iteratively reweighted optimization to separate the contributions of the different compartments. The multi-parametric MR signal analysis methods are similar to fingerprinting, as one seeks the best representation of the data based on a dictionary containing the predefined signal components. \cite{tang2018multicompartment} developed a method that improves the accuracy of the multi-compartment fingerprinting with reweighted $\ell_1$ norm regularization to distinguish the contribution of each simulated water environment. \cite{duarte2020greedy} reformulated the multi-compartment fingerprinting into the non-convex constrained least squares problem. Finally, two machine learning-based methodologies have been introduced recently. The first approach provides an unsupervised technique that learns a low-dimensional data representation \citep{slator2021data}. The method assumes that a low number of the so-called canonical spectral components are present in the mean spectrum of the data. The voxel-wise spectra are then estimated as a combination of canonical spectra. The second technique by \cite{golbabaee2022off} introduces an \textit{off-the-grid} method that does not rely on a predefined dictionary but iteratively reconstructs each compartment present in the data. The approach uses an artificial neural network to compute the Bloch responses of each compartment.

In this work, we propose a new approach called the Multi-Compartment Simple Harmonic Oscillator-based Reconstruction and Estimation (MC-SHORE) to model the diffusion-relaxation MR signal that combines the continuum modelling and a signal representation in a predefined set of basis functions. We employ the three-dimensional SHORE (3D-SHORE) representation \citep{ozarslan2009simple, zucchelli2016lies} as the diffusion kernel, allowing us to properly handle the non-Gaussian diffusion process. Our approach enables the recovery of well-accepted ensemble propagator measures demonstrating clinical potential, such as the return-to-the-origin or return-to-the-plane probabilities \citep{brusini2016ensemble,boscolo2018viability}, and the orientation distribution function (ODF) in a multi-compartment scenario, reducing the contribution of the free-water (FW) compartment. The MC-SHORE approach defines the number of atoms in a dictionary based on the order of the spherical basis, unlike commonly used samplings of the parameter space \citep{benjamini2016use,kim2017diffusion}.
The main contributions of this work can be summarized as follows:
\begin{enumerate}
\item we propose a diffusion-relaxation MR modelling framework that extends the single-compartment 3D-SHORE signal representation to the multi-compartment scenario,
\item we define the scheme to construct a representation dictionary efficiently, achieving a higher diffusion-relaxation MR signal approximation accuracy under the same number of dictionary atoms as the state-of-the-art methods employing a mono-exponential diffusion kernel,
\item we define three objective functions for the MC-SHORE  optimization by assuming: 1) the voxel-wise sparsity constraint on the solution \textit{via} the $\ell_1$ norm, 2) the sparsity of the solution in the local neighbourhood of the voxel, and 3) the sparsity of the solution in the local neighbourhood given the weights to each voxel from the neighbourhood.
\end{enumerate}

%----------------------------------------------------------------------------------------------
%----------------------------------------------------------------------------------------------
%----------------------------------------------------------------------------------------------
\section{Background}
The signal attenuation in $k-$dimensional multi-contrast MR experiment can be modelled as a multivariate Laplace transform \citep{english1991quantitative,benjamini2016use,slator2021data}
\begin{equation}
    s(\boldsymbol{\theta}) = \int \cdots \int f(\boldsymbol{\rho}) \mathcal{K}(\boldsymbol{\theta}|\boldsymbol{\rho}) d\boldsymbol{\rho}, 
    \label{eq:signal_continuum}
\end{equation}
where $s(\boldsymbol{\theta})$ is the MR signal under the experimental parameter $\boldsymbol{\theta}=[\theta_1,\ldots,\theta_k]$, $\boldsymbol{\rho}=[\rho_1,\ldots,\rho_k]$ characterises the tissue with relaxations (e.g., $T_1$, $T_2/T_2^{*}$) and/or diffusion properties such as the apparent diffusion coefficient, and $\mathcal{K}(\boldsymbol{\theta}|\boldsymbol{\rho})$ is a kernel used to model medium's features.
The Eq.~\eqref{eq:signal_continuum} can be discretized over the rectangular grid
\begin{equation}
    s(\boldsymbol{\theta}) = \sum_{l_1=1}^{N_1}\cdots \sum_{l_k=1}^{N_k} f(\rho_{l_1},\ldots,\rho_{l_k})  \prod_{l=1}^k \kappa_l(\theta_{l}|\rho_{l}),
    \label{eq:mc-basic}
\end{equation}
where $N_l$ identifies the number of discretized values of the multi-dimensional variable $\boldsymbol{\rho}$ at position $l$ and $\kappa_l(\theta_l|\rho_l)$ is a separable exponential kernel used to model $l$--th medium's feature $\rho_l$. The Eq.~\eqref{eq:mc-basic} can be written in the matrix form $\mathbf{s} = \mathbf{D}\mathbf{f}$, where $\mathbf{s}\in\mathbb{R}^M$ is the vector representing MR samples, $\mathbf{D}\in\mathbb{R}^{M\times N}$ is a dictionary of discretized kernels' values and $\mathbf{f}\in\mathbb{R}^N$ is the spectrum of MR properties \citep{benjamini2016use}.

To estimate the spectrum from Eq.~\eqref{eq:mc-basic}, the nonnegative least squares approach can be used with an optional regularization term to stabilize the solution \citep{benjamini2016use,slator2021data}
\begin{equation}
    \hat{\mathbf{f}} = \argmin_{\mathbf{f}\geqslant0}|| \mathbf{s} - \mathbf{D}\mathbf{f} ||_2^2 + \lambda R(\mathbf{f}),
    \label{eq:general-mc-twoterm}
\end{equation}
where $\lambda$ is the regularization parameter that is a trade-off between the data consistency term and the regularization function $R(\mathbf{f})$.

Considering the inversion recovery sequence \citep{hutter2018integrated}, one can define the kernels modelling $T_1\text{-ADC}$ spectrum as
\begin{equation}
    \kappa_1(TI|T_1)\kappa_2(b|D) = \left(1-2\exp{\left(-\frac{TI}{T_1}\right)}\right)\exp{\left(-bD\right)}.
    \label{eq:adc-kernel}
\end{equation}
Here, the parameter $\boldsymbol{\rho}=[T_1, D]$ with $D$ being the apparent diffusion coefficient (ADC), the acquisition setup is given by $\boldsymbol{\theta}=[TI,b]$ where $TI$ is the inversion time and $b$ is the $b$-value. To construct the dictionary $\mathbf{D}$ for the problem \eqref{eq:general-mc-twoterm} one samples the space $T_1 \times D$ and obtains $N_{T_1}$ and $N_D$ distinct values. Hence, the total number of columns of the dictionary $\mathbf{D}$ is $N=N_{T_1}N_D$, as we compute the atoms using every possible combination of $T_1$ and $D$. For the sake of this work, we will refer to this method as multi-compartment ADC (MC-ADC).

The structure of kernels \eqref{eq:adc-kernel} can be modified or extended with other parameters, like $T_2/T_2^*$, depending on the acquisition procedure \citep{benjamini2016use,kim2017diffusion,slator2021data,golbabaee2022off}. However, each additional tissue parameter increases the dictionary size geometrically.

%----------------------------------------------------------------------------------------------
%----------------------------------------------------------------------------------------------
%----------------------------------------------------------------------------------------------
\section{Methodology}
In this section, we extend the Relax-SHORE \citep{bogusz2022diffusion} that assumes the diffusion-weighted MR signal is represented by a single compartment into the multi-compartment scenario.

%----------------------------------------------------------------------------------------------
%----------------------------------------------------------------------------------------------
\subsection{Diffusion representation kernel}
To represent a diffusion part of the diffusion-relaxation MR signal, we employ the 3D-SHORE basis \citep{ozarslan2013mean,zucchelli2016lies}. This functional basis comes from the solution of the eigenvalue equation of the simple harmonic oscillator \citep{ozarslan2013mean}. The signal is represented then with the
\begin{equation}
    \phi_{nlm}(q,\mathbf{u}|\zeta) = G_{nl}(q|\zeta)Y_l^m(\mathbf{u}),
\end{equation}
where $\mathbf{q}=q\mathbf{u}$ is the wave vector, $\mathbf{u}$ is the unit direction vector, the angular part $Y_l^m(\mathbf{u})$ is represented with the spherical harmonic (SH) of degree $l$ and order $m$ and the radial part $G_{nl}(q|\zeta)$ is given by
\begin{multline}
    G_{nl}(q|\zeta) = \left(\frac{2(n-l)!}{\zeta^{3/2}\Gamma(n+3/2)}\right)^{1/2}\left(\frac{q^2}{\zeta}\right)^{l/2}\\ \times \exp{\left(\frac{-q^2}{2\zeta}\right)}L_{n-l}^{l+1/2}\left(\frac{q^2}{\zeta}\right),
    \label{eq:G_nl}
\end{multline}
where $\Gamma(\cdot)$ is the gamma function and $L_n^{(\alpha)}(\cdot)$ is the associated Laguerre polynomial.
The normalized diffusion signal $E(q,\mathbf{u})$ can be written then as the linear combination of the basis functions \citep{merlet2013continuous}
\begin{equation}
    E(q,\mathbf{u}) = \sum_{l=0 \, \mathrm{(even)}}^{L} \sum_{n=l}^{\left( L+l\right)/2} \sum_{m=-l}^{l}a_{nlm}\phi_{nlm}(q,\mathbf{u}|\zeta),
    \label{eq:shore-signal}
\end{equation}
where $L$ is the order of the basis. The  term $\phi_{000}$ refers to the mono-exponential signal decay, whereas the higher order terms present deviations from the mono-exponential decay. The number of basis functions $\phi_{nlm}$ depends on the basis order $L$, and is given as $K_L=(2L+3)(L+2)(L+4)/24$.

%----------------------------------------------------------------------------------------------
%----------------------------------------------------------------------------------------------
\subsection{Dictionary generation}
Given the inversion recovery sequence and diffusion weighting experiment, we define the kernel as
\begin{equation}
    d_{nlm}(TI,q,\mathbf{u}|T_1,\hat{\zeta}) = \left(1-2\exp{\left(-\frac{\mathrm{TI}}{T_1}\right)}\right)\phi_{nlm}(q,\mathbf{u}|\hat{\zeta}),
    \label{eq:mc-shore-vector}
\end{equation}
where acquisition parameters refer to $TI$, $q$ and $\mathbf{u}$ with $q$ being related to the $b$-value as $b=4\pi^2 q^2 \tau$, effective diffusion time $\tau=\Delta-\delta/3$, $\mathbf{u}$ is the unit direction of the diffusion gradient, $\zeta$ is the scaling parameter, $\phi_{nlm}$ is a 3D-SHORE basis function \citep{ozarslan2009simple,zucchelli2016lies}. 
We can see that for each of the $N_{T_1}$ sampled $T_1$ the diffusion is represented by the set of the basis functions $\phi_{nlm}$. We will refer to the proposed method as the multi-compartment SHORE (MC-SHORE).

We construct the dictionary for the MC-SHORE method using subdictionaries $\boldsymbol{\Psi}_c$, each defined for a fixed parameter vector $\boldsymbol{\rho}_c=[(T_1)_c,\hat{\zeta}]$
\begin{equation}    \boldsymbol{\Psi}_c\hspace{-0.06cm}=\hspace{-0.06cm}\begin{bmatrix}
        d_{000}(TI_1,q_1,\mathbf{u}_1|\boldsymbol{\rho}_c) & \hspace{-0.18cm}\cdots \hspace{-0.18cm} & d_{LLL}(TI_1,q_1,\mathbf{u}_1|\boldsymbol{\rho}_c) \\
        d_{000}(TI_2,q_2,\mathbf{u}_2|\boldsymbol{\rho}_c) & \hspace{-0.18cm}\cdots\hspace{-0.18cm} & d_{LLL}(TI_2,q_2,\mathbf{u}_2|\boldsymbol{\rho}_c)  \\
        \vdots & \hspace{-0.18cm}\ddots\hspace{-0.18cm} & \vdots\\
        d_{000}(TI_M,q_M,\mathbf{u}_M|\boldsymbol{\rho}_c) & \hspace{-0.18cm}\cdots\hspace{-0.18cm} & d_{LLL}(TI_M,q_M,\mathbf{u}_M|\boldsymbol{\rho}_c)  \\
    \end{bmatrix}.
    \label{eq:sub_dictionary}
\end{equation}

\noindent
The columns of the subdictionary $\boldsymbol{\Psi}_c$ are constructed upon the basis functions $d_{nlm}(TI,q,\mathbf{u}|\boldsymbol{\rho}_c)$ computed for subsequent data samples available in the data set. The rows in matrix \eqref{eq:sub_dictionary} follows the ordering given by Eq.~\eqref{eq:shore-signal}. 
The full MC-SHORE dictionary $\mathbf{D}\in \mathbb{R}^{M\times (N_{T_1}K_L)}$ can be constructed  from the subdictionaries $\boldsymbol{\Psi}_c$ as follows
\begin{equation}
\mathbf{D}^\mathrm{full} =  \left[\boldsymbol{\Psi}_1,\boldsymbol{\Psi}_2,\ldots,\boldsymbol{\Psi}_{N_{T_1}}\right].
\label{eq:dict_D}
\end{equation}

The estimated coefficients $\mathbf{f}\in\mathbb{R}^{N_{T_1}K_L}$ are not necessarily non-negative as for the MC-ADC. Moreover, they are not normalized because we do not divide the diffusion-relaxation signal by the $b=0$ signal as is done for diffusion analysis \citep{ozarslan2013mean}. To obtain the normalized $\tilde{\mathbf{f}}$ we divide each element of the vector $\mathbf{f}$ by the estimated proton density parameter $PD$. This parameter can be computed by estimating $T_1$ spectrum for the $b=0$ data only and then sum its elements
\begin{equation}
PD = \sum_{c=1}^{N_{T_1}} (f_{b0})_c,
\label{eq:PD}
\end{equation}
where $(f_{b0})_c$ is the $c-$th element of the vector $\mathbf{f}$ estimated using only non-diffusion-weighted MR measurements.

Assuming the sparsity of the $T_1$ spectrum we can significantly reduce the computational complexity of the problem. This lower dimensional spectrum can also be used to reduce the number of atoms in our dictionary $\mathbf{D}^\mathrm{full} $. To this end, we first determine the compartments present in the non-diffusion-weighted MR signal. We can build a~set $\mathcal{C}=\left\{c: (f_{b0})_c > 0\right\}$ that consists of the indices of non-zero coefficients. Then we can build a reduced dictionary
\begin{equation}
    \mathbf{D} = \left[\boldsymbol{\Psi}_{c_1},\boldsymbol{\Psi}_{c_2},\ldots,\boldsymbol{\Psi}_{c_{\overline{\overline{\mathcal{C}}}}}\right], \ \ c_i \in \mathcal{C},
    \label{eq:mc_shore_dictred}
\end{equation}
where $\overline{\overline{\mathcal{C}}}$ is the cardinality of the set $\mathcal{C}$.

%----------------------------------------------------------------------------------------------
%----------------------------------------------------------------------------------------------
\subsection{Signal separation from tissue compartments}
The $T_1$ spectrum can be used to separate the signal contribution from different tissue compartments. We assume that the signal is composed of two water pools: intra-/extra-axonal water IEW ($T_1\leqslant1800~\mathrm{ms}$) and free-water FW ($T_1>1800~\mathrm{ms}$) \citep{nagtegaal2020fast}. To estimate the microstructural indices from multi-parametric diffusion-relaxation MR data, we follow two approaches: 
\begin{itemize}
\item \textbf{IEW+FW:} the aggregation of all representation coefficients for each detected $T_1$ value, 
\item \textbf{IEW:} the aggregation representation coefficients for each $T_1\leqslant1800~\mathrm{ms}$, which removes FW contribution.
\end{itemize}

%----------------------------------------------------------------------------------------------
%----------------------------------------------------------------------------------------------
\subsection{Objective functions}
%
%  MC-SHORE(s)
%
We propose three objective functions used to estimate $\mathbf{f}$ from sparse diffusion-relaxation MR acquisitions using the dictionary $\mathbf{D}$ defined by Eq.~\eqref{eq:mc_shore_dictred}.

\underline{In the first objective function}, we impose the sparsity constraint on the solution using the $\ell_1$ norm as the regularizer for a single voxel. The estimated $\widehat{\mathbf{f}_i}$ for $i-$th voxel can be obtained by optimizing the cost function
\begin{equation}
    \widehat{\mathbf{f}}_i = \argmin_{\mathbf{f}_i}||\mathbf{s}_i - \mathbf{D}_i\mathbf{f}_i||_2^2 + \lambda ||\mathbf{f}_i||_1,
    \label{eq:obj-single}
\end{equation}
where $\mathbf{s}_i \in \mathbb{R}^{M}$ are diffusion-weighted MR signals from $i-$th voxel and the design matrix $\mathbf{D}_i$ for $i-$th voxel is constructed for compartments detected in this voxel using the $b_0$ volumes. To construct the dictionary $\mathbf{D}_i$, we introduce a set $\mathcal{C}_i$ that is composed of compartments indices present in the $i-$th voxel 
$\mathcal{C}_i=\left\{c: (f_{b0}^i)_c > 0\right\}$. 
The dictionary $\mathbf{D}_i$ is then build upon the set of parameters 
$\left\{\boldsymbol{\rho}_{c}^i \right\}_{c=1}^{\mathcal{C}_i}$.
The solution \eqref{eq:obj-single} is sparse voxel-wisely and the sparsity is controlled by the parameter $\lambda > 0$. We refer to this approach as \textbf{MC-SHORE(s)}. 

For example, we can construct a dictionary for $N_{T_1}=50$ logarithmically distributed $T_1$ values in range $T_1\in[10,5000]~\mathrm{ms}$ nad estimate the $T_1$ spectrum for $i-$th voxel. Let us assume now that only two elements of the vector $\mathbf{f}_{b0}$ are positive, e.g., the elements on $34-$th and $42-$th positions. Based on that, we can construct the diffusion-relaxometry dictionary \eqref{eq:mc_shore_dictred} using the 3D-SHORE basis under radial order $L=6$ as follows
\begin{equation}
    \mathbf{D}_i = \left[\boldsymbol{\Psi}_{34},\boldsymbol{\Psi}_{42}\right].
    \label{eq:obj-single-reduced}    
\end{equation}
The reduced dictionary given by Eq.~\eqref{eq:obj-single-reduced} has $100$ dictionary atoms, as the number of basis function equals $K_6=50$.

%
%  MC-SHORE(l)
%
\underline{The second objective function} includes spatial information, assuming that the solution could be sparse in the local neighbourhood of the voxel. We assumed a neighbourhood of the $i-$th voxel as the cube with the size of $3\times 3 \times 3$ or $5\times 5 \times 5$ voxels with the $i-$th voxel in the center of such cube. We also assumed the the local neighbourhoods do not share voxels with each other. The estimate for the local group of voxels $\hat{\mathbf{f}}_v$ can be obtained by solving the joint sparsity objective function
\begin{equation}
    \widehat{\mathbf{f}}_v = \argmin_{\mathbf{f}_v}||\mathbf{s}_v - \mathbf{D}_v'\mathbf{f}_v||_2^2 + \lambda ||\mathbf{f}_v||_1,
    \label{eq:obj-local}
\end{equation}
where $\mathbf{s}_v \in \mathbb{R}^{MV}$ is the stacked vector of diffusion-weighted MR signals from $V$ voxels from the neighborhood of the $i-$th voxel, $\mathbf{f}_v \in \mathbb{R}^{NV}$ is the stacked vector of signal representation coefficients, $\mathbf{D}_v'$ is the extended dictionary that is constructed as the Kronecker product $\mathbf{D}_v' = \mathbf{I}_V \otimes \mathbf{D}_v$ with identity matrix $\mathbf{I}_V\in\mathbb{R}^{V\times V}$.
The dictionary $\mathbf{D}_v$ is composed of the atoms with $T_1$ values present in all the $V$ voxels 
in the neighbourhood of the voxel $i$. Following the convention defined for the MC-SHORE(s), here we introduce a set $\mathcal{C}_v$ that is composed of the indices of the compartments present in the local group of voxels $v$, $\mathcal{C}_v=\bigcup_{j=1}^V \mathcal{C}_j$. Finally, the dictionary $\mathbf{D}_v$ is based on the set of parameters 
$\left\{\boldsymbol{\rho}_{c}^v \right\}_{c=1}^{\mathcal{C}_v}$. 
Henceforth, we refer to this approach as \textbf{MC-SHORE(l)}.

%
%  MC-SHORE(wl)
%
\underline{The third objective function} is the extension of the MC-SHORE(l) with the additional regularizer in the form of fused Lasso penalty \citep{tibshirani2005sparsity}. We include the weighting matrix $\mathbf{W}_v'$ that incorporates the similarities between the voxels from a local neighbourhood measured using a dot product
$w_{i, j} = \mathbf{s}_i^T \mathbf{s}_j/\left(||\mathbf{s}_i||_2||\mathbf{s}_j||_2 \right)$. The solution can be obtained by solving the following cost function
\begin{equation}
        \widehat{\mathbf{f}}_v = \argmin_{\mathbf{f}_v}||\mathbf{s}_v - \mathbf{D}_v'\mathbf{f}_v||_2^2 + \lambda_1 ||\mathbf{f}_v||_1 + \lambda_2 ||\mathbf{W}_v'\mathbf{f}_v||_1.
    \label{eq:obj-wl}
\end{equation}
where $\mathbf{s}_v \in \mathbb{R}^{MV}$ is again the stacked vector of diffusion-weighted MR signals, $\mathbf{f}_v \in \mathbb{R}^{NV}$ is the stacked vector of signal representation coefficients, $\mathbf{D}_v'$ is the extended dictionary that is constructed with the Kronecker product $\mathbf{D}_v' = \mathbf{I}_V \otimes \mathbf{D}_v$, the extended weight matrix defined as the Kronecker product $\mathbf{W}_v' = \mathbf{W}_v \otimes \mathbf{I}_V$ and $\mathbf{W}_v$ being the weighting matrix
\begin{equation}
    \mathbf{W}_v = \begin{bmatrix}
        1 & -\widetilde{w}_{0,1} & \cdots & -\widetilde{w}_{0,V-1}\\
        -\widetilde{w}_{1,0} & 1& \cdots & -\widetilde{w}_{1,V-1}\\
        \vdots & &\ddots &\vdots\\
        -\widetilde{w}_{V-1,0} & -\widetilde{w}_{V-1,1} & \cdots & 1\\
    \end{bmatrix}.
\end{equation}
The weights $w_{i,k}$ are normalized so that each column elements sum up to zero, i.e., $\widetilde{w}_{i,k} = w_{i,k}/\sum_{k\neq i} w_{i,k}$. We call this method \textbf{MC-SHORE(wl)}.

The previously introduced MC-SHORE approaches can be considered as special cases of the MC-SHORE(wl) method. If one fixes $\lambda_2=0$, it obtains the MC-SHORE(l). Further, if the local neighbourhood of the $i-$th voxel has only one element, that is, the $i-$th voxel itself, thus $V=1$, and consequently we obtain MC-SHORE(s).

%----------------------------------------------------------------------------------------------
%----------------------------------------------------------------------------------------------
\subsection{Optimization}
\label{sec:methodology:optimization}
To solve the proposed objectives, we use the alternating direction method of multipliers (ADMM) method \citep{boyd2011distributed}. The ADMM separates the original problem into smaller subproblems by variable splitting. Each subproblem is solved alternately by fixing the variables of remaining steps. We use an in-home implementation of the ADMM method in Python programming language (v. 3.8) and NumPy library (v.~1.23).

\begin{algorithm}[!t]
\caption{MC-SHORE(wl) fitting procedure}
\begin{algorithmic}[1]
\small
    \Require{$\mathbf{s}_v$, $\mathbf{D}_v$, $\boldsymbol{\lambda}_1$, $\boldsymbol{\lambda}_2$, $P$, $\alpha$, $\beta$, $\varepsilon_\mathrm{abs}$, $\varepsilon_\mathrm{rel}$  } %$\mu$, $\eta$, $k_\mathrm{conv}$,
    \Ensure{$\mathbf{f}_v$}
    \State Calculate the regularizations $\lambda_1, \lambda_2 :=\mathrm{GCV}(\boldsymbol{\lambda}_1, \boldsymbol{\lambda}_2,P)$
    \NoNumber{}
\Repeat
    \State Update $\mathbf{f}_v$: $\mathbf{f}_v^{k+1} := \left(\mathbf{D}'^T\mathbf{D}' + \alpha \mathbf{I} + \beta\mathbf{W}_v'^T\mathbf{W}_v'\right)^{-1}$
    \NoNumber{\hspace{1cm}$\times\left(\mathbf{D}'^T\mathbf{s}_v + \alpha(\mathbf{g}_v^k - \mathbf{y}_v^k) + \beta\mathbf{W}_v'(\mathbf{h}_v^k - \mathbf{z}_v^k)\right)$}
    \State Update $\mathbf{g}_v$: $\mathbf{g}_v^{k+1} := S_{\lambda_1/\alpha}(\mathbf{f}_v^{k+1}+\mathbf{y}_v^k)$
    \State Update $\mathbf{h}_v$: $\mathbf{h}_v^{k+1} := S_{\lambda_2/\beta}(\mathbf{W}_v'\mathbf{f}_v^{k+1}+\mathbf{z}_v^k)$
    \State Update $\mathbf{y}_v$: $\mathbf{y}_v^{k+1} := \mathbf{y}_v^k + \mathbf{f}_v^{k+1} - \mathbf{g}_v^{k+1}$
    \State Update $\mathbf{z}_v$: $\mathbf{z}_v^{k+1} := \mathbf{z}_v^k + \mathbf{W}_v'\mathbf{f}_v^{k+1} - \mathbf{h}_v^{k+1}$
    \NoNumber{}
    \State Compute primal residuals:
    \NoNumber{$\mathbf{r}_1^{k+1} := \mathbf{f}_v^{k+1} - \mathbf{g}_v^{k+1}$}
    \NoNumber{$\mathbf{r}_2^{k+1} := \mathbf{W}_v'\mathbf{f}_v^{k+1} - \mathbf{h}_v^{k+1}$}
    \NoNumber{}
    \State Compute dual residuals:
    \NoNumber{$\mathbf{s}_1^{k+1} := -\alpha (\mathbf{g}_v^{k+1}-\mathbf{g}_v^k)$}
    \NoNumber{$\mathbf{s}_2^{k+1} := -\beta (\mathbf{h}_v^{k+1}-\mathbf{h}_v^k)$}
    \NoNumber{}
    \State Compute the tolerances:
    \NoNumber{$\varepsilon_{\mathrm{pri}} := \varepsilon_{\mathrm{abs}}\sqrt{NV}$}
    \NoNumber{$\hspace{1cm}+ \varepsilon_{\mathrm{rel}} \max \left\{\norm{\mathbf{W}_v'\mathbf{f}_v^k}_2, \norm{\mathbf{f}_v^k}_2, \norm{\mathbf{g}_v^k}_2,\norm{\mathbf{h}_v^k}_2\right\}$}
    \NoNumber{$\varepsilon_{\mathrm{dual}} := \varepsilon_{\mathrm{abs}}\sqrt{NV} + \varepsilon_{\mathrm{rel}} \max \{\norm{\mathbf{y}_v^k}_2,\norm{\mathbf{z}_v^k}_2$\}}
    \NoNumber{}
    \State $k:= k+1$
    \Until{ \begin{tabular}{@{\hspace*{0.5em}}l@{}}
        $\max\left\{\norm{\mathbf{r}_1^k}_2,\norm{\mathbf{r}_2^k}_2\right\} > \varepsilon_{\mathrm{pri}}$ \\
        \Or $\max\left\{\norm{\mathbf{s}_1^k}_2,\norm{\mathbf{s}_2^k}_2\right\} > \varepsilon_{\mathrm{dual}}$ \\
      \end{tabular}  }
\end{algorithmic}
\label{alg:mc-shore-wlocal}
\end{algorithm}

The optimization scheme presented here relates to the general objective function defined in Eq.~\eqref{eq:obj-wl}.
First, we apply the variable splitting on Eq.~\eqref{eq:obj-wl}
\begin{multline}
    \mathcal{L} = ||\mathbf{s}_v - \mathbf{D}_v'\mathbf{f}_v||_2^2 + \lambda_1 ||\mathbf{g}_v||_1
    + \lambda_2 ||\mathbf{h}_v||_1 \\ \mathrm{s.t.} \quad \mathbf{f}_v = \mathbf{g}_v, \mathbf{W}_v'\mathbf{f}_v = \mathbf{h}_v.
    \label{eq:mc-shore-wlocal-vs}
\end{multline}
The problem \eqref{eq:mc-shore-wlocal-vs} can be relaxed and its augmented Lagrangian takes the form
\begin{multline}
    \mathcal{L}_{\alpha, \beta} = \frac{1}{2}\norm{\mathbf{s}_v - \mathbf{D}_v'\mathbf{f}_v}_2^2 + \lambda_1 \norm{\mathbf{g}_v}_1 + \lambda_2 \norm{\mathbf{h}_v}_1 \\ +
    \mathbf{y}_v^T(\mathbf{f}_v - \mathbf{g}_v) + \frac{\alpha}{2}\norm{\mathbf{f}_v - \mathbf{g}_v}_2^2 \\ + \mathbf{z}_v^T(\mathbf{W}_v'\mathbf{f}_v - \mathbf{h}_v) + \frac{\beta}{2}\norm{\mathbf{W}_v'\mathbf{f}_v - \mathbf{h}_v}_2^2,
    \label{eq:mc-shore-wlocal-al}
\end{multline}
where $\mathbf{y}_v$ and $\mathbf{z}_v$ are the Lagrangian multipliers. The solution does not depend on $\alpha$ and $\beta$, but these parameters affect algorithm convergence.
The update rules for each variable are defined as follows:
\begin{subequations}
    \begin{multline}
        \mathbf{f}_v^{k+1} = \left(\mathbf{D}'^T\mathbf{D}' + \alpha\mathbf{I} + \beta\mathbf{W}_v'^T\mathbf{W}_v'\right)^{-1}\\
           \times\left(\mathbf{D}'^T\mathbf{s}_v + \alpha (\mathbf{g}_v^k - \mathbf{y}_v^k) + \beta\mathbf{W}_v'(\mathbf{h}_v^k - \mathbf{z}_v^k)\right),
    \end{multline}
    \begin{align}
        \mathbf{g}_v^{k+1} &= S_{\lambda_1/\alpha}(\mathbf{f}_v^{k+1}+\mathbf{y}_v^k),\\
        \mathbf{h}_v^{k+1} &= S_{\lambda_2/\beta}(\mathbf{W}_v'\mathbf{f}_v^{k+1}+\mathbf{z}_v^k),\\
        \mathbf{y}_v^{k+1} &= \mathbf{y}_v^k + \mathbf{f}_v^{k+1} - \mathbf{g}_v^{k+1},\\
        \mathbf{z}_v^{k+1} &= \mathbf{z}_v^k + \mathbf{W}_v'\mathbf{f}_v^{k+1} - \mathbf{h}_v^{k+1},
    \end{align}
\end{subequations}
where $S_{\lambda/\delta}$ is a soft-thresholding operator. 
The algorithm progresses until the stopping criterion is fulfilled. \cite{boyd2011distributed} has suggested the optimization algorithm should stop if the so-called primal $\mathbf{r}^k$ and dual $\mathbf{s}^k$ residuals, defined as
\begin{subequations}
    \begin{align}
        \mathbf{r}_1^{k+1} &= \mathbf{f}_v^{k+1} - \mathbf{g}_v^{k+1}, \ \ \mathbf{r}_2^{k+1} = \mathbf{W}_v'\mathbf{f}_v^{k+1},\\
        \mathbf{s}_1^{k+1} &= -\alpha (\mathbf{g}_v^{k+1}-\mathbf{g}_v^k), \ \ \mathbf{s}_2^{k+1} = -\beta (\mathbf{h}_v^{k+1}-\mathbf{h}_v^k)
    \end{align}
\end{subequations}

\noindent
are lower than the tolerances $\varepsilon_{\mathrm{pri}}$ and $\varepsilon_{\mathrm{dual}}$, defined as 
\begin{subequations}
    \begin{multline}
        \varepsilon_{\mathrm{pri}} = \sqrt{NV}\varepsilon_{\mathrm{abs}}\\ + \varepsilon_{\mathrm{rel}} \max \{\norm{\mathbf{W}_v'\mathbf{f}_v^k}_2, \norm{\mathbf{f}_v^k}_2,\norm{\mathbf{g}_v^k}_2, \norm{\mathbf{h}_v^k}_2\},
    \end{multline}
    \begin{equation}
        \varepsilon_{\mathrm{dual}} = \sqrt{NV}\varepsilon_{\mathrm{abs}} + \varepsilon_{\mathrm{rel}} \max \{\norm{\mathbf{y}_v^k}_2,\norm{\mathbf{z}_v^k}_2\},
    \end{equation}
\end{subequations}
where $\varepsilon_{\mathrm{abs}}$ and $\varepsilon_{\mathrm{rel}}$ are the user-defined absolute and relative tolerances.
We can see that the tolerances do not depend on the size of the problem because of the normalizing factor $\sqrt{NV}$. The choice of the stopping criterion depends on the scale of the variables in the problem \citep{boyd2011distributed}. The summary of the general method for MC-SHORE is presented in Algorithm~\ref{alg:mc-shore-wlocal}.

%----------------------------------------------------------------------------------
%---------------------------------------------------------------------------------
\subsection{Hyperparameters selection}
\label{sec:methodology:hyperparameters}
The hyperparameters $\lambda_1,\lambda_2$ in Eq. \eqref{eq:obj-wl} are chosen using the Generalized Cross-Validation (GCV) method \citep{craven1978smoothing}. To this end, we define two sets of possible regularization parameters $\boldsymbol{\lambda}_1$ and $\boldsymbol{\lambda}_2$ and then construct the set of all possible combinations using the Cartesian product $\boldsymbol{\lambda} = \boldsymbol{\lambda}_1 \times \boldsymbol{\lambda}_2$.

The measurement set (448 volumes) is randomly divided into $P$ parts. For each $p-$th part, the coefficient vector $\mathbf{f}_v$ is estimated using each pair $(\lambda_1, \lambda_2)\in\boldsymbol{\lambda}$. The signal is approximated then for the remaining measurements using the obtained solution $\mathbf{f}_v$, and the sum of squared error (SSE) is calculated. The pair for which the SSE achieves the minimal value is chosen as the optimal regularization setup $\boldsymbol{\lambda}^p = (\lambda^p_1, \lambda^p_2)$ for the $p-$th subset of measurements. Replicating the procedure for each subset of measurements, the mean value across $\lambda_1^p$ and $\lambda_2^p$ is considered as the GCV-optimized regularization. In the case of MC-SHORE(s) and MC-SHORE(l), we assume $\boldsymbol{\lambda}=\boldsymbol{\lambda}_1$.

%----------------------------------------------------------------------------------------------
%----------------------------------------------------------------------------------------------
%----------------------------------------------------------------------------------------------
\section{Materials and methods}
\label{sec:materials_methods}
This section presents the in silico data generation scheme and in vivo data set used to evaluate the proposal, state-of-the-art techniques employed and their experimental setups.

%----------------------------------------------------------------------------------------------
%----------------------------------------------------------------------------------------------
\subsection{Data}
\label{sec:materials_methods:data}

%----------------------------------------------------------------------------------
%---------------------------------------------------------------------------------
\subsubsection{In vivo data}
\label{sec:materials_methods:invivo}
We use the diffusion-relaxation MR data made available by the MICCAI MultiDimensional (MUDI) 2019 Challenge organizers \citep{pizzolato2020acquiring}. The data set includes the brain scans of five healthy volunteers (2F/4M, aged 19-46). The data was acquired using Philips Achieva 3.0T MR scanner (Philips Healthcare, Best, Netherlands) using 32-channels receiver coil and ZEBRA protocol \citep{hutter2018integrated} at three different echo times $TE\in\{80,105,130\}~\mathrm{ms}$, 28 inversion times $TI\in[20,7322]~\mathrm{ms}$, and the repetition time $TR=7.5~\mathrm{s}$. Single data set includes 84 volumes acquired with no diffusion weighting and diffusion-weighted encoded volumes at four $b$-values $b\in (500,1000,2000,3000)~\mathrm{s\cdot mm^{2}}$ and $(7, 21, 35, 42)$ gradient directions per each combination $TI \times TE$, respectively. Other acquisition parameters: gradient pulse separation time/duration time: $\Delta/\delta=39.1/24.1~\mathrm{ms}$, voxel size: $2.5~\mathrm{mm}^3$, field of view: $220~\mathrm{mm}\times 230~\mathrm{mm}\times 140~\mathrm{mm}$. In total, 1344 volumes were acquired for each volunteer. For our experiments, we use only the data at $TE=80~\mathrm{ms}$ (i.e., 448 volumes out of 1344 available for each acquisition).

%---------------------------------------------------------------------------------
\subsubsection{In vivo data preprocessing and intra-subject registration}
\label{sec:materials_methods:registration}
The data sets were preprocessed, including noise removal in the complex space \citep{cordero2019complex}, reconstructed, and corrected for susceptibility-induced distortions using \texttt{topup} procedure from the FSL 5.0.11 (Analysis Group, FMRIB, Oxford, UK; \cite{jenkinson2012fsl}), and skull stripped using the FSL \texttt{bet}. The volumes were then registered in the following way. First, the volumes acquired for the same diffusion gradient direction but different $TE$ and $TI$ were registered using the volume acquired for the lowest $TE$ and the highest $TI$ as the reference volume. Second, all remaining volumes were registered based on the mutual information index \citep{viola1997alignment} using the DIPY package \citep{garyfallidis2014dipy}. The reference volume registrations were then propagated over all other volumes.

%---------------------------------------------------------------------------------
\subsubsection{In vivo data regions of interest retrieval}
The regions of interest (ROI) were computed under two forms: 1) the John Hopkins University (JHU) DTI-based white matter atlas \citep{mori2006mri} and white matter (WM) and gray matter (GM) areas from anatomical standard space T1-weighted MNI152 template \citep{grabner2006symmetric}. Specifically, we computed the fractional anisotropy (FA) parameter from a DTI estimated at $TI=3709.1 \ \mathrm{ms}$ and $TE=80 \ \mathrm{ms}$ and then warped the estimated FA to the Montreal Neurological Institute (MNI) space. We used a non-linear registration procedure (\texttt{fnirt} \textit{via} the FSL) preceded by a linear \texttt{flirt} initialization with seven degrees-of-freedom, correlation ratio cost function and spline-based interpolation \citep{jenkinson2001global, jenkinson2002improved}. The labels were warped from the MNI space to the subjects' native spaces using nearest-neighbour interpolation. We retrieved the following ROIs from the JHU-based atlas: the genu of corpus callosum (GCC), posterior limb of internal capsule (PLIC), arteria corona radiata (ACR), superior corona radiata (SCR), and superior longitudinal fasiculus (SLF) regions. We also segmented WM and GM regions from the MNI152 structural template using the FSL \texttt{fast} tool. All labels were finally warped to the subjects' native spaces using the nearest-neighbour interpolation.

%---------------------------------------------------------------------------------
\subsubsection{In silico data model}
\label{sec:materials_methods:insilico_model}
The parameters of \textit{in silico} data (i.e., $b$-values, $TE$, $TI$, the number of gradients per shell) are the same as previously presented for \textit{in vivo} data. The diffusion model consists of three components: intracellular water representing restricted diffusion, extracellular water presenting hindered diffusion and free water. Restricted diffusion is modeled using the Watson distribution ($\kappa=0.3$) of impermeable sticks with the longitudinal diffusivity $D_\parallel=1.5\times 10^{-3}~\mathrm{mm^2\cdot s^{-1}}$ parallel to the stick orientation. Hindered diffusion is modeled as the Watson distribution ($\kappa=0.3$) of axisymmetric tensors characterized by diffusivities parallel $D_\parallel=1.5\times 10^{-3}~\mathrm{mm^2\cdot s^{-1}}$ and perpendicular $D_\perp=0.5\times 10^{-3}~\mathrm{mm^2\cdot s^{-1}}$ to the main eigenvector of the tensor. The free diffusion component follows a mono-exponential signal with $D_\mathrm{iso}=3.0\times 10^{-3}~\mathrm{mm^2\cdot s^{-1}}$. Each stick population has $25\%$ contribution of the non-free-water signal and tensor population has the $25\%$ contribution in the non-free-water signal. For instance, if $f_\mathrm{iso}=0.2$, each stick population has $0.2$ and each tensor population $0.2$ contributions to the signal. We also assume the signal is composed of two $T_1$ values, namely the intra-/extra-axonal water (IEW) compartment is characterized by $T_1^{\mathrm{IEW}}=1000~\mathrm{ms}$ and the FW compartment by $T_1^{\mathrm{FW}}=2000~\mathrm{ms}$ \citep{nagtegaal2020fast}. The proton density $PD$ has been fixed to $PD=100$.

\subsubsection{In silico data generation}
\label{sec:materials_methods:insilico_generation}
The FW volume fraction $f_\mathrm{iso}$ varies in the interval $[0,1]$ with the step of $0.1$. We assume the presence of two fibers populations crossing at different angles $\measuredangle \in [10^\circ,90^\circ]$ with the step of $10^\circ$. Each fibre has the same diffusion properties. The Rician noise is added to the generated signal to obtain the data at different signal-to-noise ratios (SNRs). The signal at $b=0$ and the highest inversion time $TI=7322.7~\mathrm{ms}$ has been considered as the reference in computing the SNR. We generate 100 noisy data instances for each SNR.

\begin{figure*}[!t]
    \centering
    \includegraphics[width=1.0\textwidth]{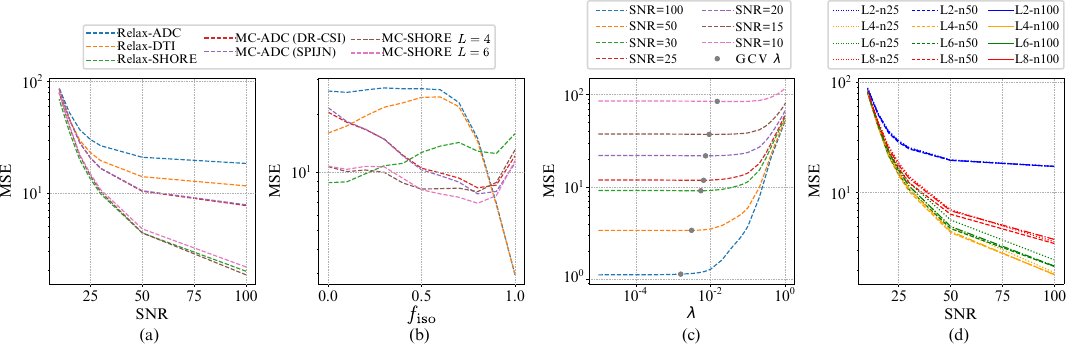}
    \caption{(a) The mean-squared error (MSE) of approximated \textit{in silico} diffusion-relaxation MR signal using the Relax-ADC, Relax-DTI, and Relax-SHORE \citep{bogusz2022diffusion}, MC-ADC method optimized with DR-CSI \citep{kim2017diffusion} and SPIJN \citep{nagtegaal2020fast}, and the proposed MC-SHORE(s) technique  as a function of (a) the signal-to-noise ratio (SNR) under a fixed $f_\mathrm{iso}=0.2$, (b) free-water contribution $f_\mathrm{iso}$ under fixed $\mathrm{SNR}=30$, (c) regularization $\lambda$ for various SNRs, and (d) dictionary size of MC-SHORE method expressed in terms of radial order $L$ and the number of $T_1$ values. Dots in panel (c) refer to the values obtained using the generalized cross-validation (GCV) under $f_\mathrm{iso}=0.2$. Panels (a), (b) and (d) present the averaged results over all possible fibre crossing configurations (angles) and noise instances. The experiments have been repeated over 100 times for each configuration setup (i.e., crossing fibers, $f_\mathrm{iso}$, SNR).  }
    \label{fig:silico-snr-fw}
\end{figure*}

%----------------------------------------------------------------------------------
%---------------------------------------------------------------------------------
\subsection{Methods}
\label{sec:materials_methods:methods}
For comparative purposes, we use the Relax-ADC \citep{hutter2018integrated,bogusz2022diffusion}, Relax-DTI \citep{de2016resolving,bogusz2022diffusion} and Relax-SHORE methods \citep{bogusz2022diffusion}, and two continuum modelling approaches, namely diffusion-relaxation correlation spectrum imaging (DR-CSI) \citep{kim2017diffusion} and sparsity promoting iterative joint non-negative least squares (SPIJN) \citep{nagtegaal2020fast}. Both continuum modelling methods employ an ADC-based kernel \eqref{eq:adc-kernel}. We also compared the results with the Relax-SHORE \citep{bogusz2022diffusion}. The method uses the 3D-SHORE basis to represent the diffusion in the multi-parametric acquisitions and enables the estimation of some microstructural measures: the return-to-the-origin, return-to-the-axis, return-to-the-plane probabilities (RTOP, RTAP, RTPP) and mean-squared displacement \citep{ozarslan2013mean,zucchelli2016lies}. Besides, we calculate the generalized fractional anisotropy (GFA) from the ODF representation \citep{tuch2004q}.

The specific tunable parameters of the methods are summarized below:
\begin{itemize}
\item \textbf{Relax-SHORE} \citep{bogusz2022diffusion}: The radial order equals $L = 6$, scale parameter $\zeta = 1/(8 \pi^2\tau \mathrm{MD})$ with the MD being the mean diffusivity estimated from a DTI at $b\leqslant 1000~\mathrm{s\cdot mm^{-2}}$.
\item \textbf{MC-ADC:} The dictionary $\mathbf{D}$ uses $N_{T_1}=50$ geometrically sampled $T_1$ values ($T_1\in [10,5000]~\mathrm{ms}$) and $N_D=50$ geometrically sampled apparent diffusion coefficients ($D \in [10^{-4},10^{-2}]~\mathrm{mm^2\cdot s^{-1}}$). The total number of atoms in the MC-ADC is 2500. In both variants, i.e., \textbf{DR-CSI} \citep{kim2017diffusion} and \textbf{SPIJN} \citep{nagtegaal2020fast}, the regularization has been fixed to $\lambda=10^{-2}$.
\item \textbf{MC-SHORE:} The radial order $L \in \{2, 4, 6, 8\}$, the scale $\zeta = 1/(8 \pi^2\tau \mathrm{MD})$. The dictionary $\mathbf{D}$ is build upon $N_{T_1}=50$ geometrically sampled $T_1$ values ($T_1\in [10,5000]~\mathrm{ms}$) and the coefficients of 3D-SHORE representation at different radial orders $L=(2, 4, 6, 8)$ producing respectively (7, 22, 50, 95) basis functions for characterising a single compartment. 
The total number of atoms is $22 \times 50 = 1100$ ($50\times50= 2500$) for the 4th (6th) order of the 3D-SHORE basis. The atoms with no contribution to the signal have been removed to reduce the computational complexity. The neighbourhoods for the  MC-SHORE(l) have been set to $3\times 3 \times 3$ and $5\times 5 \times 5$ voxels and for MC-SHORE(wl) $3\times 3 \times 3$ voxels. The absolute and relative tolerances in the stopping criterion are $\varepsilon_{\text{abs}}=10^{-4}$ and $\varepsilon_{\text{rel}}=10^{-5}$. The initial $T_1$ spectrum is estimated with the SPIJN approach \citep{nagtegaal2020fast} at $b=0$. The regularization $\lambda=1$ in this initial step is chosen so that the myelin water component ($T_1<200\,\text{ms}$) is absent in the signal fitting. The nonzero coefficients are then used to construct the reduced dictionary \eqref{eq:mc_shore_dictred}. We used fixed penalty parameters values $\alpha=10^{-5}, \beta=10^{-5}$ chosen empirically to reduce the number of ADMM iterations. The setup for the GCV-based regularization has been selected as follows: $\boldsymbol{\lambda}_1=\boldsymbol{\lambda}_2=\{10^{-3}, 10^{-2}, 10^{-1}, 10^0\}$ and the number of parts into which the measurement set is divided is $P=5$.

\end{itemize}

\begin{figure*}[!t]
    \centering
    \includegraphics[width=0.9\textwidth]{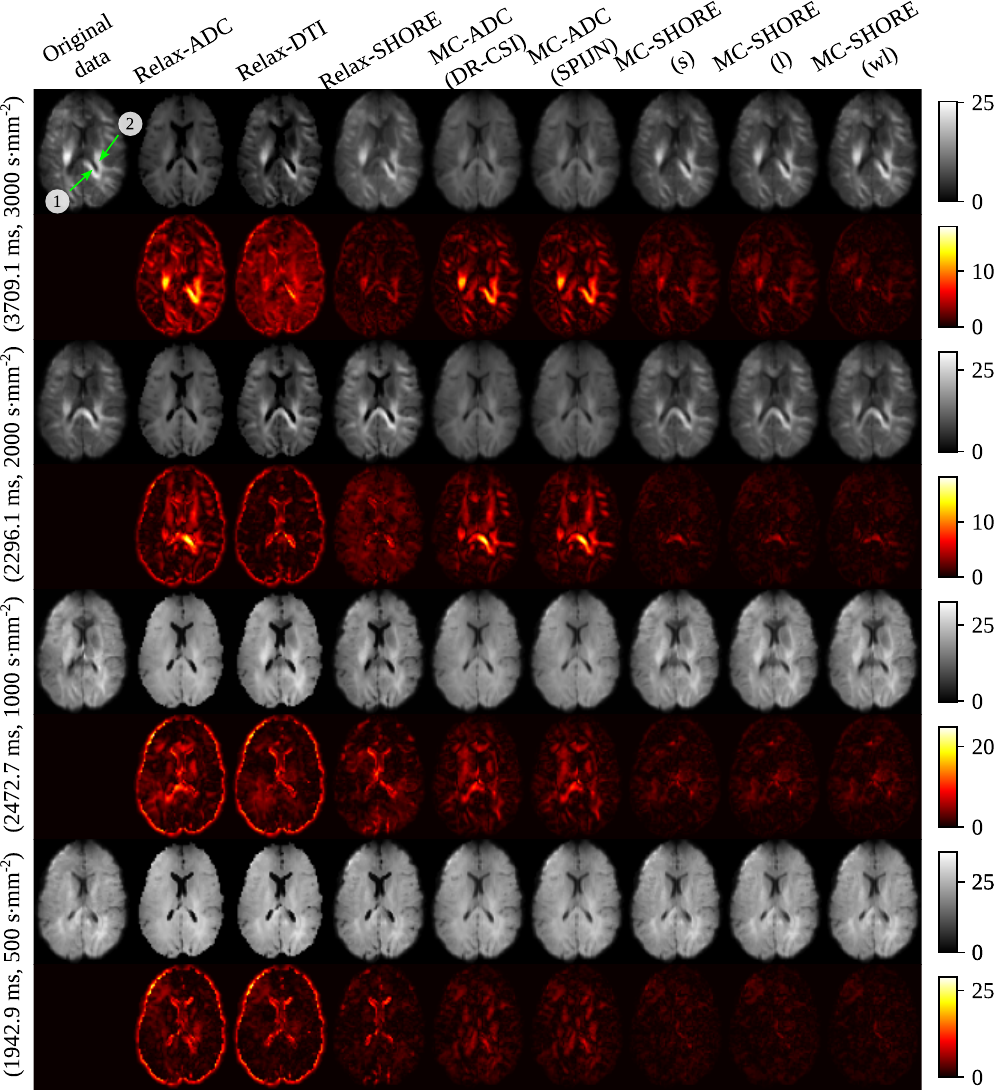}
    \caption{The approximated \textit{in vivo}  diffusion-relaxation MR data of subject \texttt{cdmri0011} at axial slice 30 (top rows) and absolute bias (bottom rows) between the original data and the approximations obtained using Relax-ADC, Relax-DTI, Relax-SHORE, MC-ADC (DR-CSI), MC-ADC (SPIJN), MC-SHORE(s), MC-SHORE(l) ($5\times5\times5$) and MC-SHORE(wl). All 448 volumes at echo time $TE = 80 \ \text{ms}$ have been used to represent the data. The approximations and absolute bias images are presented for four selected acquisitions defined in terms of ($TI$, $b$-value). Green arrows points to the splenium of the corpus callosum region (Region 1) and to the posterior limb of internal capsule (Region~2). }
    \label{fig:vivo-err-visual}
\end{figure*}

\begin{table*}[!t]
\renewcommand{\arraystretch}{1.15}
\caption{The mean-squared error (MSE) of approximated \textit{in vivo} diffusion-relaxation MR data using Relax-ADC, Relax-DTI, Relax-SHORE, MC-ADC (DR-CSI), MC-ADC (SPIJN), and the proposed MC-SHORE method under three variants, i.e., MC-SHORE(s), MC-SHORE(l), and MC-SHORE(wl). All 448 volumes from each subject at echo time $TE=80 \ \mathrm{ms}$ were used to approximate the data. The single column presents the MSE calculated for the acquisition setup specified by the inversion time ($TI$) and $b$-value. The MSE has been computed in the white matter (WM) and gray matter (GM) regions, and then averaged across five subjects. The lowest value within each acquisition setup and brain region is given in \textbf{bold}.}
\footnotesize
\begin{tabular}{ p{110pt} p{35pt} p{35pt} p{35pt} p{35pt} p{35pt} p{35pt} p{35pt} p{35pt} }
\toprule
 & \multicolumn{2}{c}{Acquisition I}&\multicolumn{2}{c}{Acquisition II} & \multicolumn{2}{c}{Acquisition III} & \multicolumn{2}{c}{Acquisition IV}\\
  & \multicolumn{2}{c}{$(3709.1, 3000)$}&\multicolumn{2}{c}{$(2296.1, 2000)$} & \multicolumn{2}{c}{$(2472.7, 1000)$} & \multicolumn{2}{c}{$(1942.9, 500)$}\\
 \cmidrule(lr){2-3}\cmidrule(lr){4-5}\cmidrule(lr){6-7}\cmidrule(lr){8-9}
 & WM& GM& WM& GM& WM& GM& WM& GM\\
 \midrule
Relax-ADC&8.276 &3.538 &6.315 &1.616 &8.368 &3.057 &4.571 &2.307 \\
Relax-DTI&5.270 &4.149 &1.459 &1.673 &4.721 &2.551 &1.728 &1.857 \\
Relax-SHORE&1.550 &1.453 &1.293 &1.376 &1.472 &2.560 &1.233 &1.751 \\ \hline
MC-ADC (DR-CSI) &5.926 &1.042 &5.715 &0.819 &6.380 &1.438 &3.975 &1.576 \\
MC-ADC (SPIJN) &5.908 &1.049 &5.643 &0.811 &6.319 &1.354 &3.738  &1.382 \\ \hline
MC-SHORE(s) &1.676 &0.726 &0.920 &0.481 &\textbf{1.389} &1.151 &1.100 &0.862 \\
MC-SHORE(l) ($3\times 3\times 3$) &1.629 &0.729 &0.929 &0.463 &1.428 &1.179 &1.048 & 0.822\\
MC-SHORE(l) ($5\times 5\times 5$) &1.623 &0.721 &0.931 &0.451 &1.434 &\textbf{1.145} &1.090 &0.843 \\
MC-SHORE(wl) ($3\times 3\times 3$) &\textbf{1.067} &\textbf{0.509} &\textbf{0.816} &\textbf{0.412} &1.444 &1.160 &\textbf{0.951} &\textbf{0.742} \\
\bottomrule
\end{tabular}
\label{tab:vivo-mse}
\begin{minipage}{20cm}
\vspace{0.02cm}
{\raggedright The acquisition parameters refer to the inversion time [$\text{ms}$] and $b$-value [$\mathrm{s}\cdot\text{mm}^{-2}$]), respectively.}
\end{minipage}
\label{tab:mse_results}
\end{table*}

\begin{figure*}[!t]
    \centering
    \includegraphics[width=0.75\textwidth]{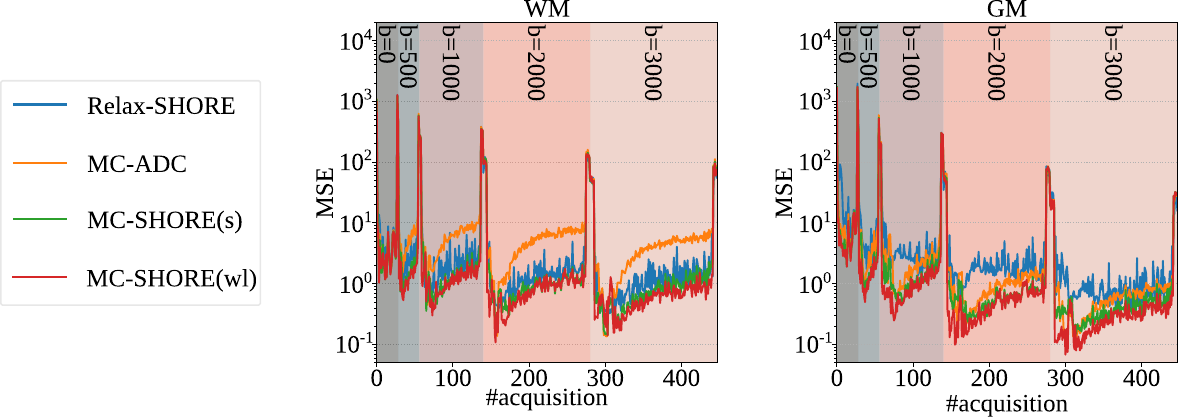}
    \caption{The MSE of approximated \textit{in vivo} diffusion-relaxation MR signal as a function of volume number (\#acquisition) given for the white matter (WM) and gray matter (GM) areas. All 448 volumes from subject \texttt{cdmri0011} have been used to approximate the data. Each plot is categorized into five blocks according to $b$-values available in the data set, including (28, 28, 84, 140, 168) acquisitions at $b=(0, 500, 100, 2000, 3000) \ \mathrm{s}\cdot\text{mm}^{-2}$, respectively. The acquisitions within each block have been sorted non-decreasingly according to the inversion time.}
    \label{fig:vivo-mse-acq}
\end{figure*}

%----------------------------------------------------------------------------------------------
%----------------------------------------------------------------------------------------------
%----------------------------------------------------------------------------------------------
\section{Experimental results}
\label{sec:experiments}

%----------------------------------------------------------------------------------
%---------------------------------------------------------------------------------
\subsection{In silico experiments: signal approximation}
\label{sec:experiments:insilico}
This very first experiment considers the possibility of the diffusion-relaxation signal to be approximated using the proposed method. We relate one compartment methods Relax-ADC \citep{hutter2018integrated}, Relax-DTI \citep{de2016resolving}, Relax-SHORE \citep{bogusz2022diffusion}, MC-ADC kernel under the DR-CSI \citep{kim2017diffusion} and SPIJN \citep{nagtegaal2020fast} optimizations with the MC-SHORE(s) optimized \textit{via} the objective function \eqref{eq:obj-single}. The experiments inspect how accurately, in terms of mean squared error (MSE), the methods approximate synthetically generated diffusion-relaxation signal under a range of SNRs and free-water volume fraction $f_\mathrm{iso}$, and additionally. Additionally, the experiment investigates the effects of the regularization parameter $\lambda$ and dictionary size on the MC-SHORE(s). The results are presented in Fig.~\ref{fig:silico-snr-fw}. The reference used in this experiment is the noise-free diffusion-relaxation signal, i.e., the $\mathrm{SNR}=\infty$.

First, regarding the MSE as a function of SNR illustrated in Fig.~\ref{fig:silico-snr-fw}(a), we observe the non-directional methods Relax-ADC and MC-ADC, and the diffusion tensor-based Relax-DTI behave the worst for the whole range of SNR considered in the experiment. On the contrary, the directional techniques Relax-SHORE and MC-SHORE show significantly better behaviour in terms of MSE. Here, the free-water volume fraction in the synthetic diffusion-relaxation signal has been fixed to $f_\mathrm{iso}=0.2$. The single- (Relax-SHORE) and multi-compartment (MC-SHORE) representations do not show any relevant differences between each other. This experiment also illustrates the importance of delivering the least noisy diffusion-relaxation signal possible for data approximation, i.e., the MSE reduces by one order of magnitude from a typical SNR observed in clinical diffusion-weighted data to almost noiseless data at $\mathrm{SNR}=100$.

\begin{figure*}[!t]
    \centering
    \includegraphics[width=1.0\linewidth]{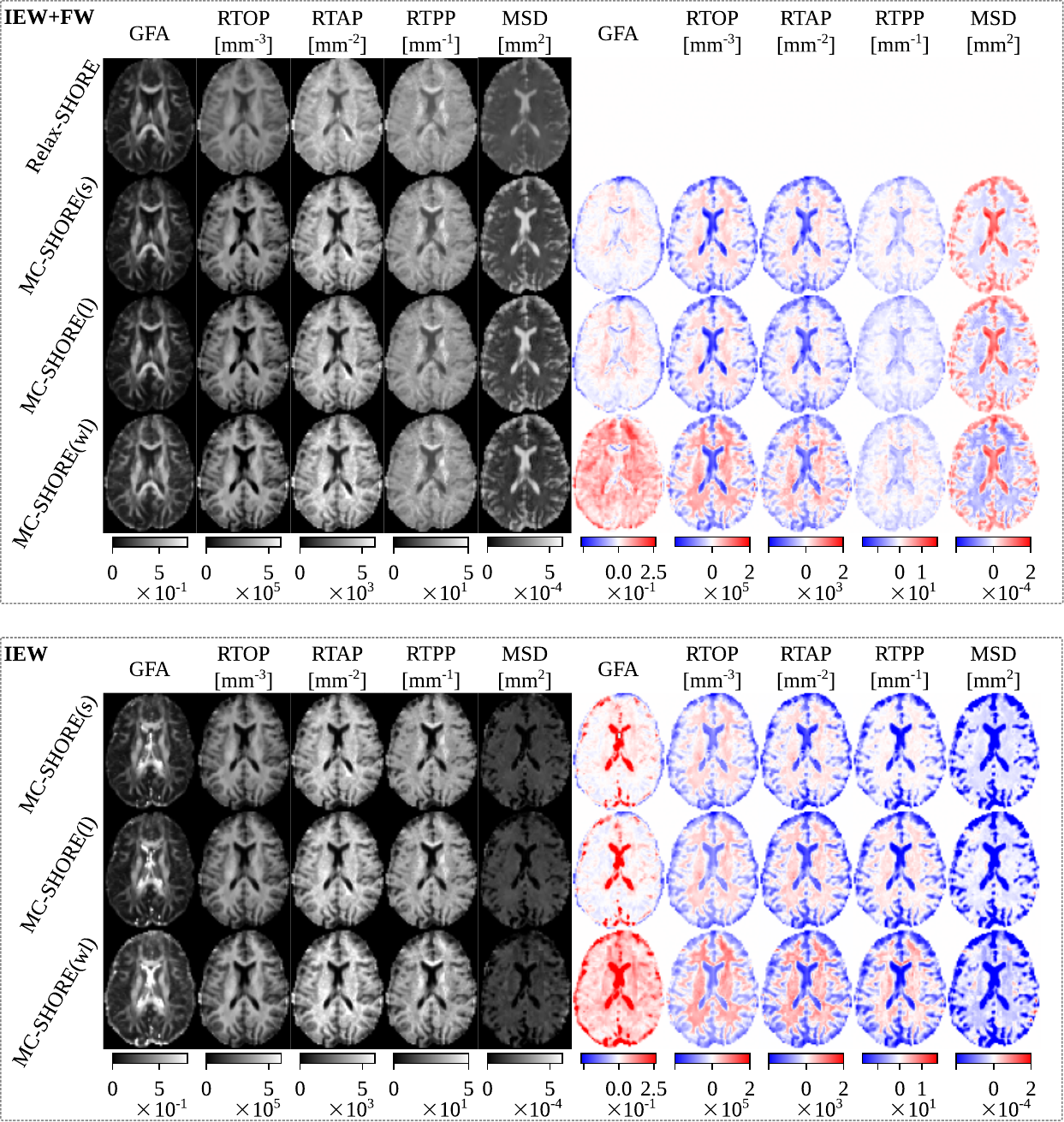}
    \caption{Estimated microstructural indices from the \textit{in vivo} diffusion-relaxation MR data (subject \texttt{cdmri0011}, slice 30) using single-compartment Relax-SHORE and the proposed multi-compartmental MC-SHORE approach under tree variants, i.e., MC-SHORE(s), MC-SHORE(l) and MC-SHORE(wl). The top figure presents the measures calculated under the assumption of intra-axonal, extra-axonal and free water compartments (IEW+FW), while the bottom considers only intra- and extra-axonal compartments (IEW). The figures in the right panels illustrate absolute bias between the indices calculated using the single-compartment Relax-SHORE and particular variants of multi-compartmental MC-SHORE. }
    \label{fig:vivo-measures}
\end{figure*}

\begin{figure*}[!t]
    \centering
    \includegraphics[width=\textwidth]{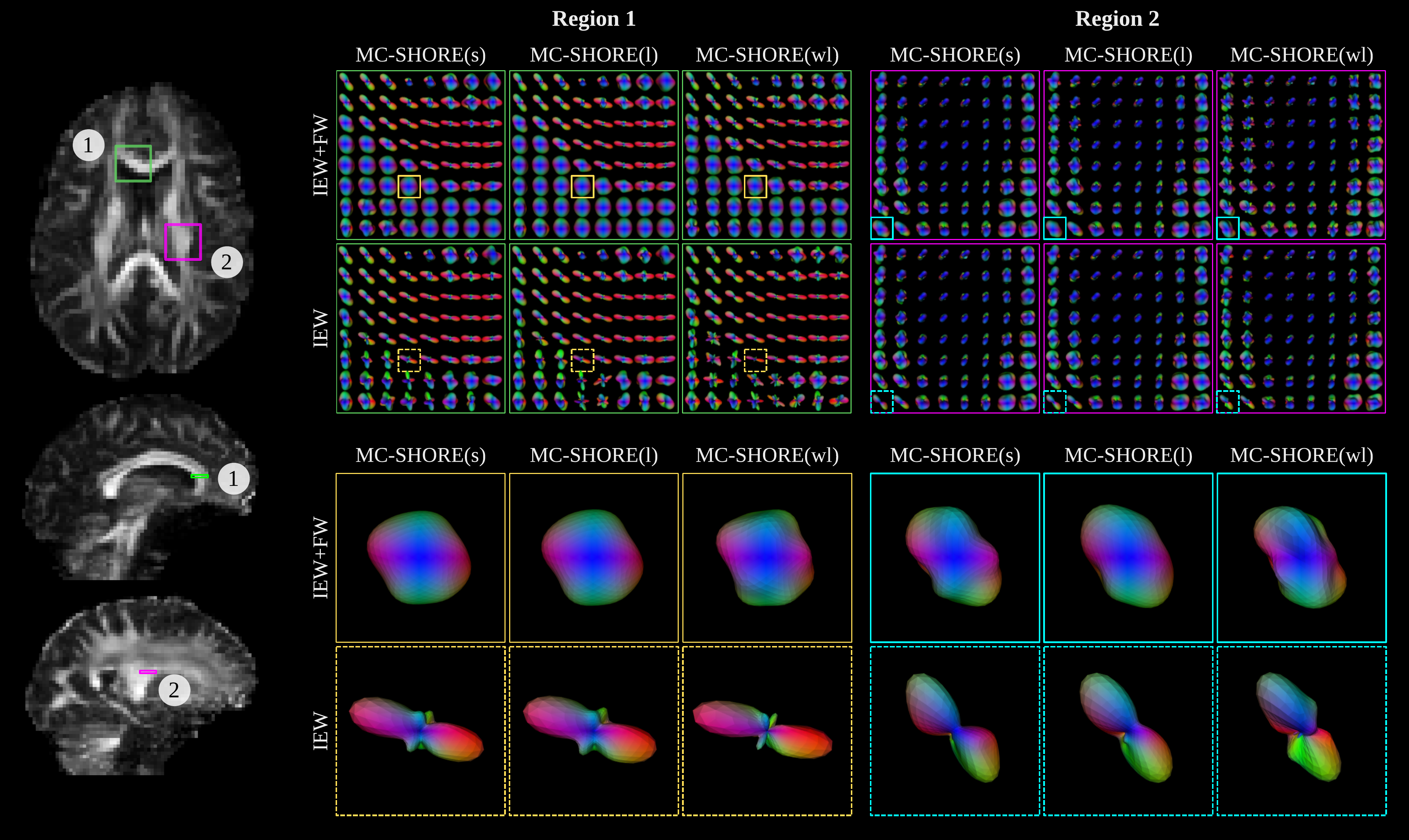}
    \caption{The orientation distribution functions (ODFs) estimated under the assumption of intra-axonal, extra-axonal and free water compartments (IEW+FW) and intra-/extra-axonal compartments (IEW) using MC-SHORE(s), MC-SHORE(l), MC-SHORE(wl). The green rectangle indicates the genu of the corpus callosum region (Region 1), while the violet rectangle refers to the posterior limb of internal capsule (Region 2). Individual ODFs from all variants have been zoomed in the bottom panel.}
    \label{fig:vivo_odfs}
\end{figure*}

\begin{figure*}[!t]
     \centering
     \includegraphics[width=0.95\textwidth]{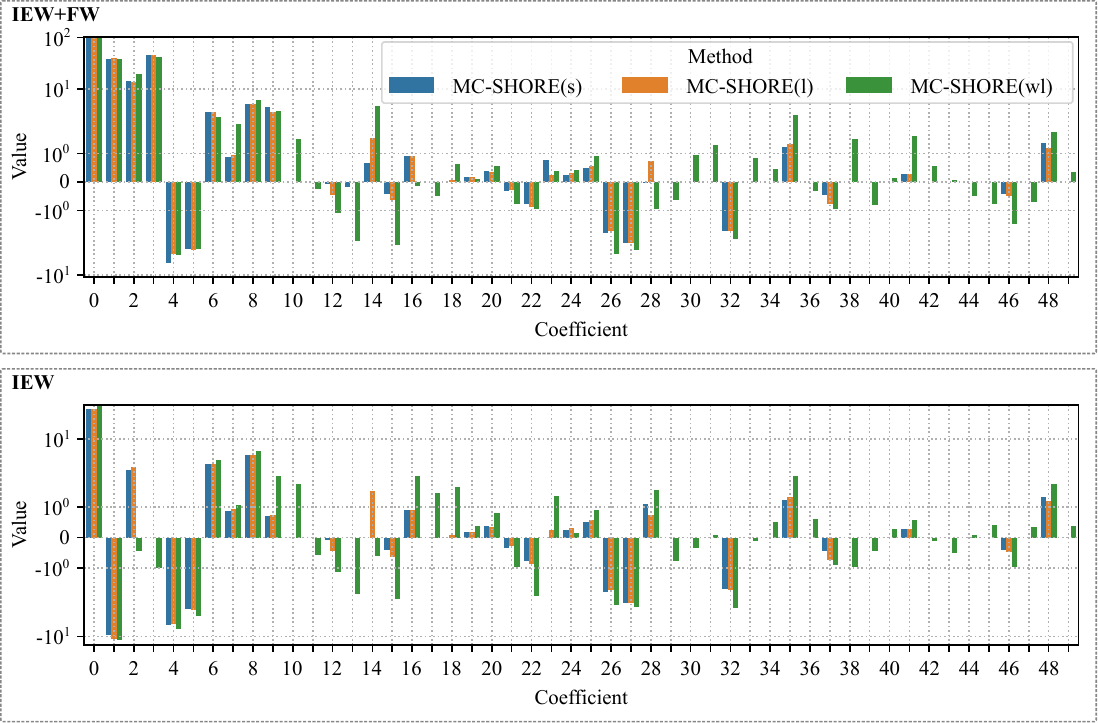}
     \caption{The coefficients $\phi_{nlm}(q,\mathbf{u}|\zeta)$ representing the diffusion-relaxation MR signal using the MC-SHORE method in the selected voxels from the GCC area (see yellow squares in Fig.~\ref{fig:vivo_odfs}). The top row presents the coefficients computed for intra-axonal, extra-axonal and free water compartments (IEW+FW), while the bottom shows the intra-/extra-axonal compartments (IEW) variant. All coefficients have been divided by the proton density $\mathrm{PD}$ as given in Eq.~\eqref{eq:PD}.}
     \label{fig:vivo-coefs}
\end{figure*}

The second experiment illustrates the MSE as a function of free-water volume fraction $f_\mathrm{iso}$ and a fixed $\mathrm{SNR}=30$ (see Fig.~\ref{fig:silico-snr-fw}(b)). We consider the MC-SHORE at two different radial orders, i.e., $L=4$ and $L=6$, which show comparable behaviours for the whole range of $f_\mathrm{iso}$. Not surprisingly, the Relax-ADC and Relax-DTI decrease their MSE for a very high $f_\mathrm{iso}$, showing a much smaller approximation error than the proposal. However, as recently demonstrated by \cite{pieciak2023spherical}, such high free-water values (i.e., $f_\mathrm{iso}>0.9$) appear very rarely in the brain's WM. Considering a real-world scenario ($f_\mathrm{iso}<0.9$), the MC-SHORE under both radial order configurations outperforms MC-ADC techniques and mostly the Relax-SHORE. Note that the MC-SHORE presents quite the opposite behaviour to the Relax-SHORE, i.e., the higher the free-water volume fraction, the smaller the MSE. These results might be due to a prevailing multi-compartmental nature of the signal under higher free-water values.

\begin{figure*}[!t]
    \centering
    \includegraphics[width=1.0\linewidth]{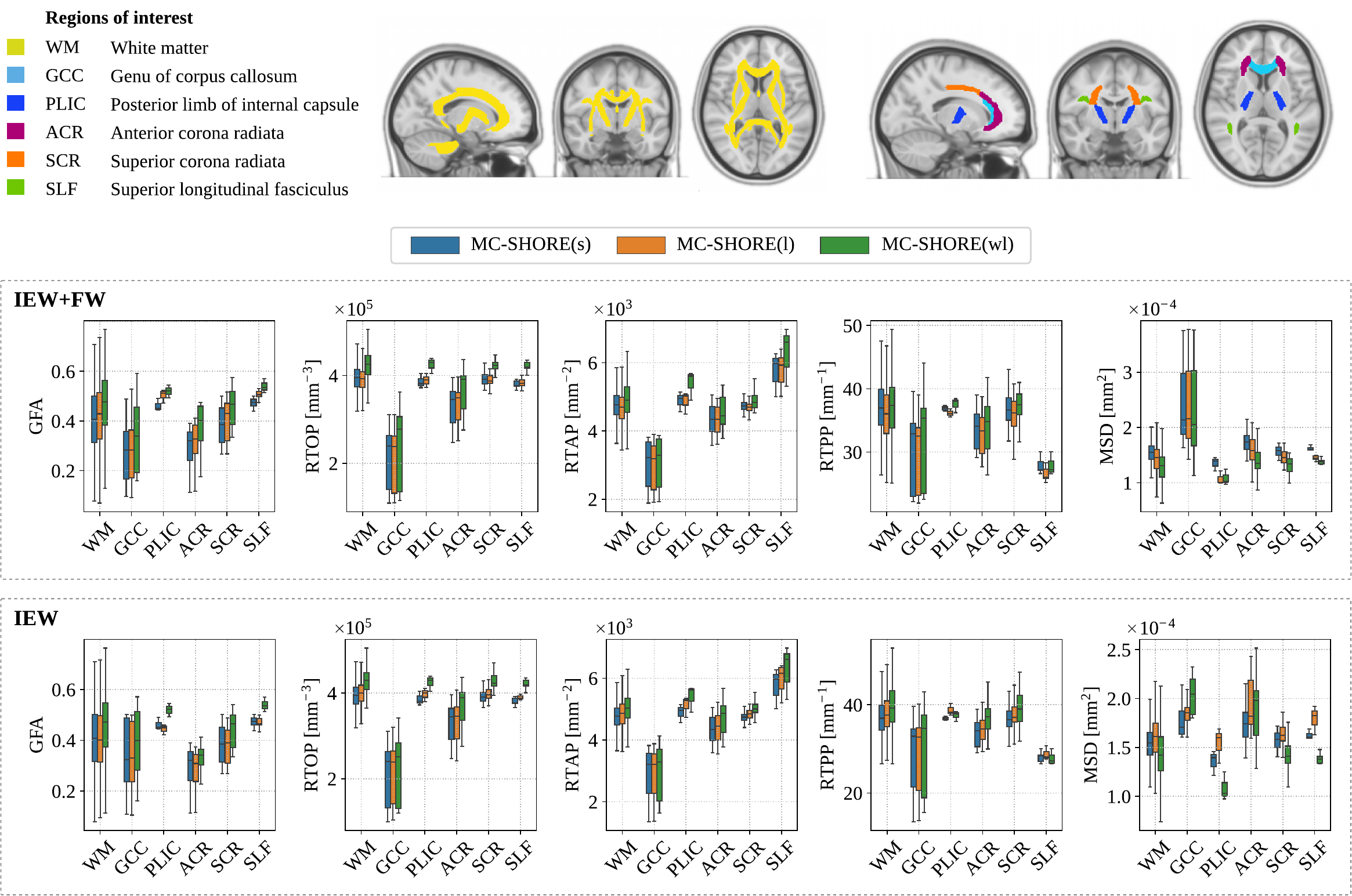}
    \caption{Boxplots present the statistics of microstructural indices GFA, RTOP, RTAP, RTPP, and MSD for subject \texttt{cdmri0013} over the selected regions of interest. The measures were estimated using MC-SHORE(s), MC-SHORE(l) and MC-SHORE(wl) from all available samples under $TE=80 \ \mathrm{ms}$ (448 volumes). The top set of boxplots presents the measures calculated under the assumption of intra-axonal, extra-axonal and free-water compartments (IEW+FW), while bottom boxplots show the measures obtained given intra- and extra-axonal water compartments (IEW). Boxplots are characterized by the first quartile $Q_1$, median and third quartile $Q_3$, while the whiskers denote $Q_1 - 1.5\times \mathrm{IQR}$ and $Q_3 + 1.5\times \mathrm{IQR}$ with $\mathrm{IQR}$ being the inter-quartile range.}
    \label{fig:vivo-regions-0013}
\end{figure*}

The third experimental result, depicted in Fig.~\ref{fig:silico-snr-fw}(c), inspects the dependence of approximation MSE on the selection of the regularization parameter $\lambda$ used in the MC-SHORE(s) method under different SNR levels. Here, we use the GCV method to automatically select the optimal $\lambda$ for each scenario and compare such derived MSE to the MSE obtained for a range of reasonable $\lambda$ values. We observe the increase in the MSE once $\lambda$ exceeds the value of $10^{-2}$. All in all, the preferable choice of $\lambda$ is rather small, smaller than $10^{-2}$, and the GCV method fulfils its role correctly.

Eventually, in the last \textit{in silico} experiment, we inspect the MC-SHORE(s) under different dictionary sizes expressed in terms of radial order $L \in \{ 2, 4, 6, 8 \}$ and the number of $T_1$ values, $T_1 \in \{ 25, 50, 100 \}$. The results are presented in Fig.~\ref{fig:silico-snr-fw}(d). We can observe that the highest MSE is present for the radial order of $L=2$ for all tested SNR. On the contrary, the lowest MSE is reported for the radial order of 4 and 6 with slightly higher values in the case of $L=6$. We note that selecting higher radial orders, e.g. $L=8$, increases MSE, perhaps due to model overfitting. The number of $T_1$ values has a minor effect on the MSE.

%----------------------------------------------------------------------------------
%---------------------------------------------------------------------------------
\subsection{In vivo experiments: signal approximation}
\label{sec:experiments:invivo}
The second group of experiments focus on signal approximation using \textit{in vivo} diffusion-relaxation MR data. Here, we use all the previously-mentioned state-of-the-art methods and the proposed MC-SHORE under the radial order $L=6$ and three variants, namely MC-SHORE(s), MC-SHORE(l) and  MC-SHORE(wl).

\begin{figure*}[!t]
    \centering
    \includegraphics[width=1.0\textwidth]{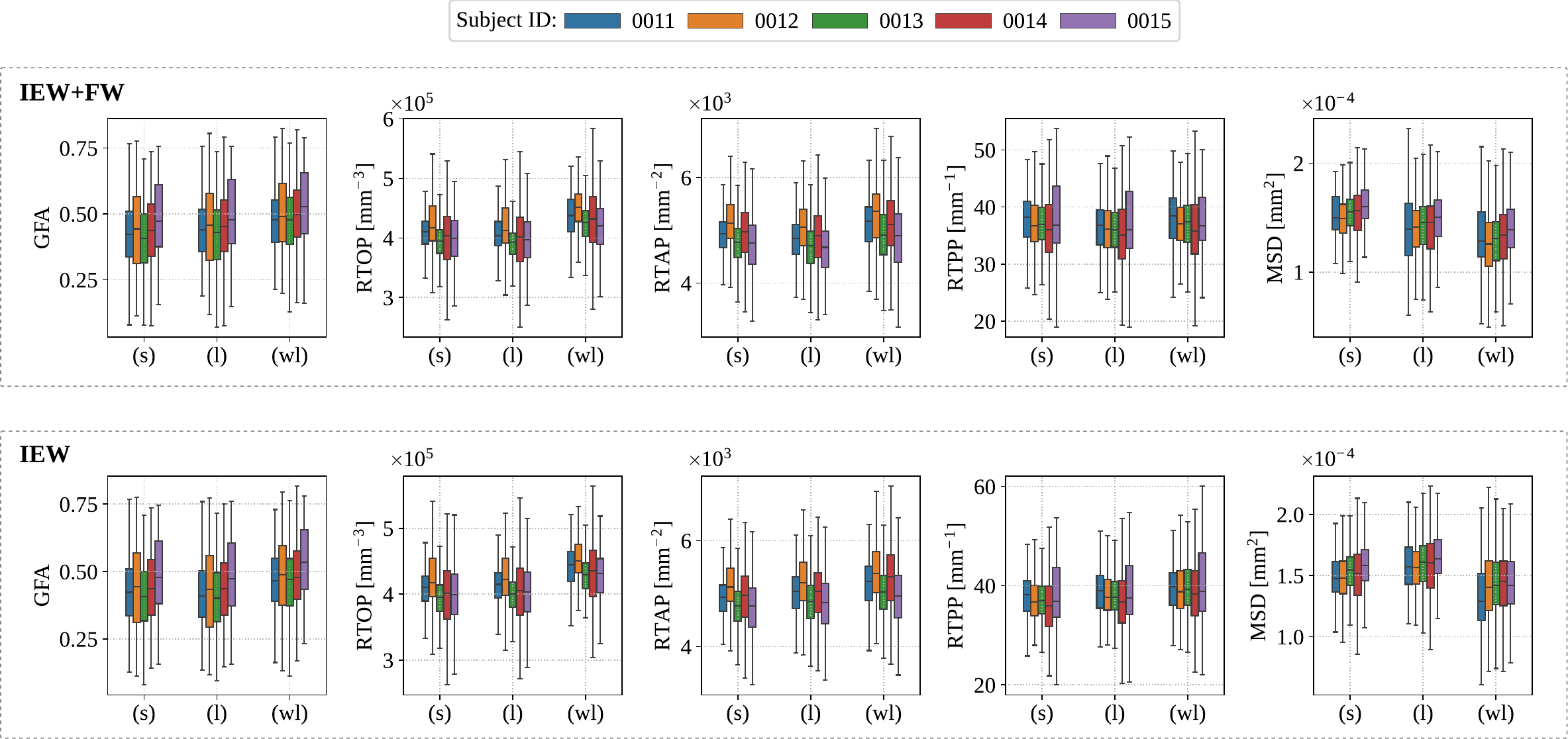}
    \caption{Boxplots present the statistics of microstructural indices GFA, RTOP, RTAP, RTPP, and MSD over white matter for all five subjects considered in the study. The measures were estimated using MC-SHORE(s), MC-SHORE(l) and MC-SHORE(wl) from all available samples under $TE=80 \ \mathrm{ms}$ (448 volumes). The top set of boxplots presents the measures calculated under the assumption of intra-axonal, extra-axonal and free-water compartments (IEW+FW), while bottom boxplots show the measures obtained given intra- and extra-axonal water compartments (IEW). Boxplots are characterized by the first quartile $Q_1$, median and third quartile $Q_3$, while the whiskers denote $Q_1 - 1.5\times \mathrm{IQR}$ and $Q_3 + 1.5\times \mathrm{IQR}$ with $\mathrm{IQR}$ being the inter-quartile range.}
    \label{fig:vivo-regions-all}
\end{figure*}

We start with the visual results for subject \texttt{cdmri0011} and four selected acquisition variants given in terms of inversion time $TI$ and $b$-value that are shown in Fig.~\ref{fig:vivo-err-visual}. The top rows demonstrate the original MR signals acquired under the specified acquisition parameters and the approximations using different methods. The bottom rows show the absolute bias between the original signals and the approximations. We note while Fig.~\ref{fig:vivo-err-visual} illustrates selected four acquisitions, the approximation has been calculated using all 448 volumes available for $TE=80 \ \mathrm{ms}$. This experiment shows a relatively high absolute error of signal approximations using non-directional methods Relax-ADC and MC-ADC and tensor-based Relax-DTI, primarily in the regions of high diffusion anisotropy, i.e., in the splenium of corpus callosum (SCC) or PLIC (see the arrows drawn in Fig.~\ref{fig:vivo-err-visual}). The potential improvement in signal approximation with Relax-SHORE under lower and moderate $b$-values is not evident from visual inspections, albeit under a high $b$-value regime (i.e., $b=3000 \ \mathrm{s}\cdot\text{mm}^{-2}$) is fairly noticeable. We note the Relax-SHORE method presents the approximated diffusion-relaxation MR signal considering a single compartment exists in a voxel. The approximation precision has been significantly enhanced using the multi-compartmental MC-SHORE technique across all verified acquisition setups. This experiment does not determine which MC-SHORE variant delivers the lowest absolute bias.

In addition to the visual experiment, we present the quantitative results explained using the MSE of approximated diffusion-relaxation MR signals. We use identical data acquisition setups, as in the previous experiment presented in Fig.~\ref{fig:vivo-err-visual}, but hither, we compute the MSE over WM and GM areas across all five subjects. The aggregated results are presented in Table~\ref{tab:vivo-mse}. We observe a substantial reduction in the approximation error considering the MC-SHORE over non-directional methods, tensor-based Relax-DTI, and the single-compartment directional Relax-SHORE. This improvement is evident for the WM area, where the diffusion-weighted MR signal cannot be retrieved with a non-anisotropic scalar-based coefficient or even the second-order tensor. Regarding the variants of the MC-SHORE, we observe the advantage of the MC-SHORE(wl) over other variants, especially under a higher $b$-value regime.

In the next experiment, we look closer to the MSE of approximated \textit{in vivo} diffusion-relaxation MR signal as a function of volume number. Specifically, we approximate the data for the subject \texttt{cdmri0011} using all 448 volumes available at echo time $TE = 80 \ \mathrm{ms}$ and then compute the MSE separately for each volume starting from $b=0$ until $b=3000 \ \mathrm{s}\cdot\mathrm{s}^{-2}$. The results for the MC-SHORE(s) and MC-SHORE(wl) together with the Relax-SHORE and MC-ADC (DR-CSI) have been depicted in Fig.~\ref{fig:vivo-mse-acq}. As a side note, the MSE values illustrated in Fig.~\ref{fig:vivo-mse-acq} have been sorted over each $b$-value block non-decreasingly according to the inversion time. We can observe that MC-SHORE(wl) mostly provides the lowest approximation error over WM and GM areas compared to other techniques. The MSE computed for the approximated volumes at very low/high inversion times exhibit significant uppings.

%----------------------------------------------------------------------------------
%---------------------------------------------------------------------------------
\subsection{In vivo experiments: microstructural measures}
\label{sec:experiments:invivo_micro}
The next experiment visually inspects microstructural indices computed using the MC-SHORE under three variants (MC-SHORE(s), MC-SHORE(l) and MC-SHORE(wl)) and two scenarios (IEW+FW and IEW) and relates them to a single-compartmental Relax-SHORE method \citep{bogusz2022diffusion}. The results for subject \texttt{cdmri0011} and a selected axial slice has been included in Fig.~\ref{fig:vivo-measures}. We observe higher values of the GFA index for the MC-SHORE(wl) variant compared to the MC-SHORE(s) and MC-SHORE(l) in both scenarios, i.e., IEW+FW and IEW. The increase in GFA measure is particularly noticeable in the regions of anisotropic diffusion, such as the WM. One can localize these changes in the right panel of Fig.~\ref{fig:vivo-measures} in terms of the absolute bias between the MC-SHORE(wl) and Relax-SHORE. Other measures like the RTxP or MSD also differ from the Relax-SHORE but to a much lesser extent than the GFA. All in all, we observe some deviations in all measures computed under all variants of the MC-SHORE method. These differences show the evidence that multi-compartmental MC-SHORE deviates from the single-compartmental Relax-SHORE, and according to the previous experiments depicted in Figs.~\ref{fig:vivo-err-visual} and \ref{fig:vivo-mse-acq}, these deviations translate to a finer tissue representation using the multi-compartmental model.

%----------------------------------------------------------------------------------
%---------------------------------------------------------------------------------
\subsection{In vivo experiments: qualitative results}
\label{sec:experiments:odfs}
The following experiment visually inspects the orientation distribution functions (ODFs) estimated using the MC-SHORE(s), MC-SHORE(l) and MC-SHORE(wl) under $L=6$ (see Fig.~\ref{fig:vivo_odfs}). We consider two scenarios: 1) the tissue is composed of intra-axonal, extra-axonal and free-water compartments, and 2) only intra-/extra-axonal water compartments exist. The ODF functions indeed identify the genu of the corpus callosum (GCC) (see Region 1). However, the thickness of the GCC measured in the transverse plane appears to be more extensive in the case of IEW compared to the sum of signals coming from all compartments (IEW+FW). The IEW variant enabled the extraction of directional information where the partial volume effect is particularly prominent (see the yellow squares in the second row in Fig.~\ref{fig:vivo_odfs})). This information is concealed under the IEW+FW scenario. However, contrary in the IEW variant, we observe some spurious peaks in the estimated ODFs. A similar behaviour of the MC-SHORE method has been observed for the posterior limb of the internal capsule (PLIC) (see Region 2). However, in this case, the spurious peaks are suppressed. All in all, the IEW MC-SHORE uncovers a directional nature of the underlying fibre structure, hitherto "blurred" by the partial volume effect in the IEW+FW scenario.

The 3D-SHORE coefficients used to estimate the ODF from Region 1 in Fig.~\ref{fig:vivo_odfs} have been depicted in Fig.~\ref{fig:vivo-coefs}. Considering the differences between the IEW+FW and IEW (cf. the coefficients number 1 between IEW+FW and IEW variants; coefficients start from zero), we can observe that the FW contribution is dominant, thus removing it from the aggregated signal reveals the intra-axonal/extra-axonal water content. Comparing the coefficients across MC-SHORE variants, we can see that the MC-SHORE(wl) approach gives a less sparse solution.

%----------------------------------------------------------------------------------
%---------------------------------------------------------------------------------
\subsection{In vivo experiments: region- and population-based study}
\label{sec:experiments:population}
In the final two experiments, we demonstrate region- and population-based distributions of estimated microstructural measures using the MC-SHORE method under three variants considered previously, namely MC-SHORE(s), MC-SHORE(l) and MC-SHORE(wl) and two scenarios (i.e., IEW+FW and IEW). The measures were estimated from all samples available in the data sets at $TE=80 \ \mathrm{ms}$ (448 volumes). The regions of interest for each subject were retrieved from the JHU WM atlas using the procedure described in section~\ref{sec:materials_methods:registration}.

First, in Fig.~\ref{fig:vivo-regions-0013}, we present the distributions of the measures (GFA, RTxP, MSD) for subject \texttt{cdmri0013} across the selected regions of interest. Each distribution is represented using a boxplot characterized by the first, the second (median) and the third quartile. In general, the GFA and RTxP measures computed with MC-SHORE(wl) present increased values compared to MC-SHORE(s) and MC-SHORE(l) in both scenarios IEW+FW and IEW. This effect could have been noticed before for the GFA in Fig.~\ref{fig:vivo-measures}. The opposite behaviour is pronounced for the MSD measure, i.e., the MC-SHORE(wl) mostly supplies decreased MSD values.
As we can observe in the Fig.~\ref{fig:vivo_odfs} removing the free water contribution reduces the mean value of the ODF thus increasing the GFA index value. The RTxP indices reflects the inverse of diffusivity, the high diffusivity produces low RTxP values. By removing the FW content in the regions where the diffusion is mainly restricted or hindered the diffusivity is lower, thus the indices have higher values. In the region of CSF, that is mainly composed of the free water, removing it leaves almost nothing so the RTxP values are nullified. Removing FW content with high diffusivity also implies the decrease of the MSD.

Second, in Fig.~\ref{fig:vivo-regions-all}, we illustrate the microstructural measures calculated across all five subjects considered in this study. We limit the results to the boxplots computed from the entire WM area. In this experiment, we observe a diversified behaviour of the measures, i.e., GFA, RTOP, and RTAP are somewhat variable across the subjects, but the RTPP and MSD seem to be more consistent.

%----------------------------------------------------------------------------------
%---------------------------------------------------------------------------------
\section{Discussion}
\label{sec:discussion}
The \textit{in silico} results demonstrated in Fig.~\ref{fig:silico-snr-fw}(a) and Fig.~\ref{fig:silico-snr-fw}(b) indicate that an improvement in the approximation accuracy (explained in terms of the MSE between the approximated data and the reference) of diffusion-relaxation MR signal can be achieved by employing the spherical basis for the diffusion part of the kernel in Eq.~\eqref{eq:adc-kernel}, for instance, the 3D-SHORE basis \citep{zucchelli2016lies}. Indeed, the Relax-SHORE outperforms the Relax-ADC and Relax-DTI approaches for the whole range of tested SNR and reasonable FWVF values. These results corroborate previous arrangements made by \cite{bogusz2022diffusion} but consider the more complex \textit{in silico} tissue model defined here. In this paper, however, we introduced a new multi-compartment diffusion-relaxation MR signal representation built upon the Relax-SHORE method \citep{bogusz2022diffusion}, namely the MC-SHORE approach. The MC-SHORE methodology enables the modelling diffusion-relaxation MR signal acquired from highly undersampled multi-parametric acquisition strategies like the ZEBRA protocol \citep{hutter2018integrated}, but considering the received signal comes from many non-exchanging tissue compartments. The MC-SHORE has been proposed under three scenarios, namely MC-SHORE(s), MC-SHORE(l) and MC-SHORE(wl). The first variant imposes the sparsity constraint on the solution using the $\ell_1$ norm, the second assumes the solution is sparse in the local neighbourhood of the voxel, and the last one extends the MC-SHORE(l) with the fused Lasso penalty, enabling to introduce the weights assigned to each voxel from the neighbourhood. Predictably, the engagement of the 3D-SHORE basis has enabled catching the anisotropic features (likewise the Relax-SHORE) but assumes the multi-compartmental nature of the tissue, like the MADCO \cite{benjamini2016use} or MC-ADC \citep{kim2017diffusion,nagtegaal2020fast}. However, the MADCO and MC-ADC model the diffusion in each compartment as a monoexponential decay, thus obscuring the directional nature of the diffusion process.

The \textit{in silico} results translate directly to the \textit{in vivo} results, i.e., the MC-SHORE approach approximates the diffusion-relaxation MR signal more accurately in terms of the MSE than the multi-compartmental MC-ADC technique over the WM and GM areas. Precisely, for highly anisotropic water diffusion environments, the MC-ADC approach does not approximate the diffusion-MR signal as accurately as the MC-SHORE does in terms of signal approximation MSE (see Table \ref{tab:vivo-mse}, Figs.~\ref{fig:vivo-err-visual} and \ref{fig:vivo-mse-acq}). The increase in the MSE observed for the high FWVF contribution might be caused by the lower density of the $T_1$ sampling defined for higher $T_1$ values. This issue is absent for single-compartment methods such as the Relax-ADC or Relax-DTI \citep{hutter2018integrated,de2016resolving,bogusz2022diffusion}. We note, however, that such high values of the FWVF (i.e., $f_\text{iso} > 0.9$) exist in the WM area sporadically, as presented recently by \cite{pieciak2023spherical}.

Our proposal employs the 3D-SHORE basis as a diffusion kernel in Eq.~\eqref{eq:mc-shore-vector}. The proposal is flexible enough to use different spherical bases, such as the multiple q-shell diffusion propagator imaging (mq-DPI) \citep{descoteaux2011multiple}, Bessel Fourier orientation reconstruction (BFOR) \citep{hosseinbor2013bessel}, mean apparent propagator magnetic resonance imaging (MAP-MRI; \cite{ozarslan2013mean}) or the radial basis functions \citep{ning2015estimating,wang2024q}. However, we decided to choose the 3D-SHORE for two reasons: 1) it enables the calculation of a wide range of microstructural features from closed-form equations, including the MSD and RTxP measures \citep{zucchelli2016lies} compared to the mq-DPI and BFOR, and 2) it provides a trade-off between the detailed angular representation of the signal and the complexity given by the MAP-MRI.
All in all, engaging the spherical 3D-SHORE basis has allowed us to represent the diffusion-relaxation MR signal more accurately and estimate the ODF and a variety of tissue parameters compared to simple exponential decay representations. At the same time, the number of the atoms in the dictionary remains the same or even lower in comparison to the MADCO or MC-ADC \citep{benjamini2016use,kim2017diffusion,nagtegaal2020fast}. Recently, \cite{filipiak2022performance} proposed the ODF estimation methodology based on the large dictionaries, which may also be extended to multi-parametric imaging, as our scenario considers.

The proposed method enables the separation of various compartments characterised by different $T_1$ relaxations. The 3D-SHORE basis, compared to the simple kernel-based approaches, incorporates the information about the direction of the diffusion process and also approximates the signal more accurately in the high \textit{b}-value regime \citep{ozarslan2009simple,ozarslan2013mean,zucchelli2016lies,bogusz2022diffusion}. This information can be effectively used to estimate the fibre directions present within the imaged voxel. As presented in Fig.~\ref{fig:vivo_odfs}, the FW compartment hinders the specific intra-voxel directional information and yields the partial volume effect \citep{conturo1995diffusion,alexander2001analysis,vos2011partial}. However, the microstructural indices, especially the GFA (see Fig.~\ref{fig:vivo-measures}), estimated for the IEW variant, seem sensitive if the initial compartment identification is not precise. That indicates at least two new problems to be tackled: 1) to develop a better procedure to estimate the initial compartment distribution or 2) to derive new general multi-compartmental indices adjusted for diffusion-relaxometry MRI. Excluding the FW compartment (i.e., the IEW variant) can effectively reduce the partial volume effect mainly present in voxels on the boundary between two tissues under the IEW+FW scenario (see the bottom row in Fig.~\ref{fig:vivo_odfs}). The inter-subject indices distributions (see Fig.~\ref{fig:vivo-regions-all}) computed over the WM region indicate that the proposed method is generally stable across various datasets. The variabilities observed for RTOP and RTAP have been previously reported by \cite{bouhrara2023adult} for adult lifespan study. 

The estimated indices from the MC-SHORE slightly differ from the ones obtained using the single-compartment Relax-SHORE (see Fig.~\ref{fig:vivo-measures}). The dissimilarities are especially observable in the high CSF contribution areas for the RTxP and MSD indices. Such differences may be caused by the low accuracy of the Relax-SHORE representation in these regions, conceivably due to the fixed initial estimate of the $T_1$ parameter \citep{bogusz2022diffusion}. The incorporation of the signal similarity measure in the optimization process of MC-SHORE(wl) increases the GFA values in the WM regions compared to the MC-SHORE(s) and MC-SHORE(l) variants (see Fig.~\ref{fig:vivo-regions-0013}). The additional regularization term in the objective function of the MC-SHORE(wl) variant offers overall less sparse solutions (see Fig.~\ref{fig:vivo-coefs}). The sparsity of the differences between the coefficients in the voxels in the local neighbourhood does not impose the sparsity of the solution itself \citep{tobisch2019comparison,fick2018non}. Consequently, it increases the variance of the ODF, and thus, the GFA index also holds higher values.

The regularization included in the optimization cost function, in general, may enforce some features of the reconstructed signal, such as the non-negativity \citep{haije2020enforcing,tristan2023hydi}. The MC-SHORE method regularizes the optimization problem to reduce the number of solutions by assuming that the signal has some unique properties, such as sparse representation on a predefined basis or the similarity of the signal exists in the neighbouring voxels. The key to successful regularization lies in choosing the proper regularization parameter $\lambda$ (or parameters $\lambda_1$ and $\lambda_2$ in the case of MC-SHORE(wl)), which is not trivial. We used the GCV methodology \citep{craven1978smoothing} to automatically choose such parameters based on a predefined set of possible regularization parameters. Here, we restricted the predefined set of parameters to just four values $\boldsymbol{\lambda}_1=\boldsymbol{\lambda}_2 = \{10^{-3},10^{-2},10^{-1},10^0\}$. However, on the one hand, such a small set of possible regularization parameters might not be optimal, but on the other, increasing the set leads to increase the complexity of the optimization procedure, especially for the MC-SHORE(wl) case (i.e., linear increase in the number of elements in $\boldsymbol{\lambda}_1$ and $\boldsymbol{\lambda}_2$ leads to a squarely increase in number of elements in $\boldsymbol{\lambda}$). Recently, \cite{canales2021revisiting} proposed an approach to regularize the problem based on the Bayesian inference. Our proposal could be reformulated to use such Bayesian regularization.

%----------------------------------------------------------------------------------
%----------------------------------------------------------------------------------
\subsection{Study limitations}
The presented study has some limitations that must be noted here. First, the MC-SHORE method uses an initial estimate of the lower dimensional spectrum with \cite{nagtegaal2020fast} to reduce the number of atoms in the dictionary. This practice leads to the loss of some information related to non-detected compartments, which cannot be recovered later in the estimation procedure. Moreover, false-positive components found in this initial step affect the microstructural indices' estimation process. This is particularly noticeable in the GFA map for the IEW compartment in the lateral ventricles region (see Fig.~\ref{fig:vivo-measures}), where only the FW should be observed. 

In this work, we assumed that the myelin water contribution to the signal is negligible. The lowest TE covered in the \textit{in vivo} MUDI data is 80 ms, which is significantly higher than the $T_2$ value of myelin component \citep{lee2021so}. Nevertheless, given lower TE values, the method may be arranged to catch the myelin component.

The MC-SHORE method enables to estimate the ODFs from the 3D-SHORE representation coefficients. It has been shown that the 3D-SHORE basis restricts the angular resolution of the estimated ODFs \citep{fick2015using}. Strictly speaking, the 3D-SHORE basis may lead to troubles by incorrectly determining the crossing fibre bundles if the angle between the bundles is supposed to be less than around $45^\circ$. One can possibly improve the angular resolution using a higher order basis (parameter $L$ in Eq.~\eqref{eq:shore-signal}) or use other functional basis like MAP-MRI \citep{ozarslan2013mean}. However, as we can observe in Fig.~\ref{fig:silico-snr-fw}(d), increasing order basis beyond $L=4$ does not necessarily improve signal approximation accuracy, possibly to a relatively small number of samples considered in the estimation process for such complex model.

Next, the assumption behind the method is to discretize the parameter space, as presented in  Eq.~\eqref{eq:mc-basic}. The estimated spectrum only resembles the properties of the tissue characterized by a continuous distribution, and thus, one can expect some biases in the estimated spectrum. Nevertheless, this attribute is not strictly related to the MC-SHORE, but it is a common trait of other discretization-based methods, such as \cite{benjamini2016use}, or \cite{kim2017diffusion}.

Last but not least, the MC-SHORE employs fixed penalty parameters $\alpha, \beta$ in the ADMM algorithm. These parameters do not affect the precision of the solution but interfere with the convergence rate \citep{boyd2011distributed}. Indeed, one can use an adaptive choice of these parameters to reduce the number of iterations until the stopping criterion is satisfied.

%----------------------------------------------------------------------------------
%----------------------------------------------------------------------------------
%----------------------------------------------------------------------------------
\section{Conclusion}
This paper introduces a new multi-compartment diffusion-relaxation MR signal representation based on the 3D-SHORE functional basis. The method enables the separation of signals coming from intra-/extra-axonal and free water contributions and obtaining a variety of state-of-the-art quantitative brain measures. We have shown that our MC-SHORE solution can significantly improve the signal approximation accuracy while maintaining a reasonable number of atoms in the dictionary. The proposed approach can be further extended with other spherical representations or used to model additional features of the brain tissues in multi-parametric diffusion-relaxation imaging. 

%----------------------------------------------------------------------------------
%----------------------------------------------------------------------------------
\bibliographystyle{model4-names}
\bibliography{refs}

\begin{thebibliography}{56}
\expandafter\ifx\csname natexlab\endcsname\relax\def\natexlab#1{#1}\fi
\providecommand{\url}[1]{\texttt{#1}}
\providecommand{\href}[2]{#2}
\providecommand{\path}[1]{#1}
\providecommand{\DOIprefix}{doi:}
\providecommand{\ArXivprefix}{arXiv:}
\providecommand{\URLprefix}{URL: }
\providecommand{\Pubmedprefix}{pmid:}
\providecommand{\doi}[1]{\href{http://dx.doi.org/#1}{\path{#1}}}
\providecommand{\Pubmed}[1]{\href{pmid:#1}{\path{#1}}}
\providecommand{\bibinfo}[2]{#2}
\ifx\xfnm\undefined \def\xfnm[#1]{\unskip,\space#1}\fi
%Type = Article
\bibitem[{Afzali et~al.(2021)Afzali, Pieciak, Newman, Garyfallidis,
  {\"O}zarslan, Cheng and Jones}]{afzali2021sensitivity}
\bibinfo{author}{Afzali\xfnm[ M.]}, \bibinfo{author}{Pieciak\xfnm[ T.]},
  \bibinfo{author}{Newman\xfnm[ S.]}, \bibinfo{author}{Garyfallidis\xfnm[ E.]},
  \bibinfo{author}{{\"O}zarslan\xfnm[ E.]}, \bibinfo{author}{Cheng\xfnm[ H.]},
  \bibinfo{author}{Jones\xfnm[ D.K.]}.
\newblock \bibinfo{title}{The sensitivity of diffusion {MRI} to microstructural
  properties and experimental factors}.
\newblock \bibinfo{journal}{Journal of Neuroscience Methods}
  \bibinfo{year}{2021};\bibinfo{volume}{347}:\bibinfo{pages}{108951}.
%Type = Article
\bibitem[{Alexander et~al.(2001)Alexander, Hasan, Lazar, Tsuruda and
  Parker}]{alexander2001analysis}
\bibinfo{author}{Alexander\xfnm[ A.L.]}, \bibinfo{author}{Hasan\xfnm[ K.M.]},
  \bibinfo{author}{Lazar\xfnm[ M.]}, \bibinfo{author}{Tsuruda\xfnm[ J.S.]},
  \bibinfo{author}{Parker\xfnm[ D.L.]}.
\newblock \bibinfo{title}{Analysis of partial volume effects in
  diffusion-tensor mri}.
\newblock \bibinfo{journal}{Magnetic Resonance in Medicine: An Official Journal
  of the International Society for Magnetic Resonance in Medicine}
  \bibinfo{year}{2001};\bibinfo{volume}{45}(\bibinfo{number}{5}):\bibinfo{pages}{770--780}.
%Type = Article
\bibitem[{Basser et~al.(1994)Basser, Mattiello and LeBihan}]{basser1994mr}
\bibinfo{author}{Basser\xfnm[ P.J.]}, \bibinfo{author}{Mattiello\xfnm[ J.]},
  \bibinfo{author}{LeBihan\xfnm[ D.]}.
\newblock \bibinfo{title}{Mr diffusion tensor spectroscopy and imaging}.
\newblock \bibinfo{journal}{Biophysical journal}
  \bibinfo{year}{1994};\bibinfo{volume}{66}(\bibinfo{number}{1}):\bibinfo{pages}{259--267}.
%Type = Article
\bibitem[{Benjamini and Basser(2016)}]{benjamini2016use}
\bibinfo{author}{Benjamini\xfnm[ D.]}, \bibinfo{author}{Basser\xfnm[ P.J.]}.
\newblock \bibinfo{title}{Use of marginal distributions constrained
  optimization (madco) for accelerated 2d mri relaxometry and diffusometry}.
\newblock \bibinfo{journal}{Journal of magnetic resonance}
  \bibinfo{year}{2016};\bibinfo{volume}{271}:\bibinfo{pages}{40--45}.
%Type = Article
\bibitem[{Bogusz et~al.(2022)Bogusz, Pieciak, Afzali and
  Pizzolato}]{bogusz2022diffusion}
\bibinfo{author}{Bogusz\xfnm[ F.]}, \bibinfo{author}{Pieciak\xfnm[ T.]},
  \bibinfo{author}{Afzali\xfnm[ M.]}, \bibinfo{author}{Pizzolato\xfnm[ M.]}.
\newblock \bibinfo{title}{Diffusion-relaxation scattered mr signal
  representation in a multi-parametric sequence}.
\newblock \bibinfo{journal}{Magnetic Resonance Imaging}
  \bibinfo{year}{2022};\bibinfo{volume}{91}:\bibinfo{pages}{52--61}.
%Type = Article
\bibitem[{Boscolo~Galazzo et~al.(2018)Boscolo~Galazzo, Brusini, Obertino,
  Zucchelli, Granziera and Menegaz}]{boscolo2018viability}
\bibinfo{author}{Boscolo~Galazzo\xfnm[ I.]}, \bibinfo{author}{Brusini\xfnm[
  L.]}, \bibinfo{author}{Obertino\xfnm[ S.]}, \bibinfo{author}{Zucchelli\xfnm[
  M.]}, \bibinfo{author}{Granziera\xfnm[ C.]}, \bibinfo{author}{Menegaz\xfnm[
  G.]}.
\newblock \bibinfo{title}{On the viability of diffusion {MRI}-based
  microstructural biomarkers in ischemic stroke}.
\newblock \bibinfo{journal}{Frontiers in neuroscience}
  \bibinfo{year}{2018};\bibinfo{volume}{12}:\bibinfo{pages}{92}.
%Type = Article
\bibitem[{Bouhrara et~al.(2023)Bouhrara, Avram, Kiely, Trivedi and
  Benjamini}]{bouhrara2023adult}
\bibinfo{author}{Bouhrara\xfnm[ M.]}, \bibinfo{author}{Avram\xfnm[ A.V.]},
  \bibinfo{author}{Kiely\xfnm[ M.]}, \bibinfo{author}{Trivedi\xfnm[ A.]},
  \bibinfo{author}{Benjamini\xfnm[ D.]}.
\newblock \bibinfo{title}{Adult lifespan maturation and degeneration patterns
  in gray and white matter: A mean apparent propagator ({MAP}) {MRI} study}.
\newblock \bibinfo{journal}{Neurobiology of aging}
  \bibinfo{year}{2023};\bibinfo{volume}{124}:\bibinfo{pages}{104--116}.
%Type = Article
\bibitem[{Boyd et~al.(2011)Boyd, Parikh, Chu, Peleato, Eckstein
  et~al.}]{boyd2011distributed}
\bibinfo{author}{Boyd\xfnm[ S.]}, \bibinfo{author}{Parikh\xfnm[ N.]},
  \bibinfo{author}{Chu\xfnm[ E.]}, \bibinfo{author}{Peleato\xfnm[ B.]},
  \bibinfo{author}{Eckstein\xfnm[ J.]}, et~al.
\newblock \bibinfo{title}{Distributed optimization and statistical learning via
  the alternating direction method of multipliers}.
\newblock \bibinfo{journal}{Foundations and Trends{\textregistered} in Machine
  learning}
  \bibinfo{year}{2011};\bibinfo{volume}{3}(\bibinfo{number}{1}):\bibinfo{pages}{1--122}.
%Type = Article
\bibitem[{Brusini et~al.(2016)Brusini, Obertino, Galazzo, Zucchelli, Krueger,
  Granziera and Menegaz}]{brusini2016ensemble}
\bibinfo{author}{Brusini\xfnm[ L.]}, \bibinfo{author}{Obertino\xfnm[ S.]},
  \bibinfo{author}{Galazzo\xfnm[ I.B.]}, \bibinfo{author}{Zucchelli\xfnm[ M.]},
  \bibinfo{author}{Krueger\xfnm[ G.]}, \bibinfo{author}{Granziera\xfnm[ C.]},
  \bibinfo{author}{Menegaz\xfnm[ G.]}.
\newblock \bibinfo{title}{Ensemble average propagator-based detection of
  microstructural alterations after stroke}.
\newblock \bibinfo{journal}{International journal of computer assisted
  radiology and surgery}
  \bibinfo{year}{2016};\bibinfo{volume}{11}:\bibinfo{pages}{1585--1597}.
%Type = Article
\bibitem[{Canales-Rodr{\'\i}guez et~al.(2021)Canales-Rodr{\'\i}guez, Pizzolato,
  Yu, Piredda, Hilbert, Radua, Kober and Thiran}]{canales2021revisiting}
\bibinfo{author}{Canales-Rodr{\'\i}guez\xfnm[ E.J.]},
  \bibinfo{author}{Pizzolato\xfnm[ M.]}, \bibinfo{author}{Yu\xfnm[ T.]},
  \bibinfo{author}{Piredda\xfnm[ G.F.]}, \bibinfo{author}{Hilbert\xfnm[ T.]},
  \bibinfo{author}{Radua\xfnm[ J.]}, \bibinfo{author}{Kober\xfnm[ T.]},
  \bibinfo{author}{Thiran\xfnm[ J.P.]}.
\newblock \bibinfo{title}{Revisiting the {T}2 spectrum imaging inverse problem:
  {B}ayesian regularized non-negative least squares}.
\newblock \bibinfo{journal}{Neuroimage}
  \bibinfo{year}{2021};\bibinfo{volume}{244}:\bibinfo{pages}{118582}.
%Type = Article
\bibitem[{Conturo et~al.(1995)Conturo, McKinstry, Aronovitz and
  Neil}]{conturo1995diffusion}
\bibinfo{author}{Conturo\xfnm[ T.E.]}, \bibinfo{author}{McKinstry\xfnm[ R.C.]},
  \bibinfo{author}{Aronovitz\xfnm[ J.A.]}, \bibinfo{author}{Neil\xfnm[ J.J.]}.
\newblock \bibinfo{title}{Diffusion mri: precision, accuracy and flow effects}.
\newblock \bibinfo{journal}{NMR in Biomedicine}
  \bibinfo{year}{1995};\bibinfo{volume}{8}(\bibinfo{number}{7}):\bibinfo{pages}{307--332}.
%Type = Article
\bibitem[{Cordero-Grande et~al.(2019)Cordero-Grande, Christiaens, Hutter, Price
  and Hajnal}]{cordero2019complex}
\bibinfo{author}{Cordero-Grande\xfnm[ L.]}, \bibinfo{author}{Christiaens\xfnm[
  D.]}, \bibinfo{author}{Hutter\xfnm[ J.]}, \bibinfo{author}{Price\xfnm[
  A.N.]}, \bibinfo{author}{Hajnal\xfnm[ J.V.]}.
\newblock \bibinfo{title}{Complex diffusion-weighted image estimation via
  matrix recovery under general noise models}.
\newblock \bibinfo{journal}{Neuroimage}
  \bibinfo{year}{2019};\bibinfo{volume}{200}:\bibinfo{pages}{391--404}.
%Type = Article
\bibitem[{Craven and Wahba(1978)}]{craven1978smoothing}
\bibinfo{author}{Craven\xfnm[ P.]}, \bibinfo{author}{Wahba\xfnm[ G.]}.
\newblock \bibinfo{title}{Smoothing noisy data with spline functions:
  estimating the correct degree of smoothing by the method of generalized
  cross-validation}.
\newblock \bibinfo{journal}{Numerische mathematik}
  \bibinfo{year}{1978};\bibinfo{volume}{31}(\bibinfo{number}{4}):\bibinfo{pages}{377--403}.
%Type = Article
\bibitem[{De~Santis et~al.(2016)De~Santis, Barazany, Jones and
  Assaf}]{de2016resolving}
\bibinfo{author}{De~Santis\xfnm[ S.]}, \bibinfo{author}{Barazany\xfnm[ D.]},
  \bibinfo{author}{Jones\xfnm[ D.K.]}, \bibinfo{author}{Assaf\xfnm[ Y.]}.
\newblock \bibinfo{title}{Resolving relaxometry and diffusion properties within
  the same voxel in the presence of crossing fibres by combining inversion
  recovery and diffusion-weighted acquisitions}.
\newblock \bibinfo{journal}{Magnetic resonance in medicine}
  \bibinfo{year}{2016};\bibinfo{volume}{75}(\bibinfo{number}{1}):\bibinfo{pages}{372--380}.
%Type = Article
\bibitem[{Descoteaux et~al.(2011)Descoteaux, Deriche, Le~Bihan, Mangin and
  Poupon}]{descoteaux2011multiple}
\bibinfo{author}{Descoteaux\xfnm[ M.]}, \bibinfo{author}{Deriche\xfnm[ R.]},
  \bibinfo{author}{Le~Bihan\xfnm[ D.]}, \bibinfo{author}{Mangin\xfnm[ J.F.]},
  \bibinfo{author}{Poupon\xfnm[ C.]}.
\newblock \bibinfo{title}{Multiple q-shell diffusion propagator imaging}.
\newblock \bibinfo{journal}{Medical image analysis}
  \bibinfo{year}{2011};\bibinfo{volume}{15}(\bibinfo{number}{4}):\bibinfo{pages}{603--621}.
%Type = Article
\bibitem[{Does(2018)}]{does2018inferring}
\bibinfo{author}{Does\xfnm[ M.D.]}.
\newblock \bibinfo{title}{Inferring brain tissue composition and microstructure
  via {MR} relaxometry}.
\newblock \bibinfo{journal}{NeuroImage}
  \bibinfo{year}{2018};\bibinfo{volume}{182}:\bibinfo{pages}{136--148}.
%Type = Article
\bibitem[{Duarte et~al.(2020)Duarte, Repetti, G{\'o}mez, Davies and
  Wiaux}]{duarte2020greedy}
\bibinfo{author}{Duarte\xfnm[ R.]}, \bibinfo{author}{Repetti\xfnm[ A.]},
  \bibinfo{author}{G{\'o}mez\xfnm[ P.A.]}, \bibinfo{author}{Davies\xfnm[ M.]},
  \bibinfo{author}{Wiaux\xfnm[ Y.]}.
\newblock \bibinfo{title}{Greedy approximate projection for magnetic resonance
  fingerprinting with partial volumes}.
\newblock \bibinfo{journal}{Inverse Problems}
  \bibinfo{year}{2020};\bibinfo{volume}{36}(\bibinfo{number}{3}):\bibinfo{pages}{035015}.
%Type = Article
\bibitem[{English et~al.(1991)English, Whittall, Joy and
  Henkelman}]{english1991quantitative}
\bibinfo{author}{English\xfnm[ A.]}, \bibinfo{author}{Whittall\xfnm[ K.]},
  \bibinfo{author}{Joy\xfnm[ M.]}, \bibinfo{author}{Henkelman\xfnm[ R.]}.
\newblock \bibinfo{title}{Quantitative two-dimensional time correlation
  relaxometry}.
\newblock \bibinfo{journal}{Magnetic resonance in medicine}
  \bibinfo{year}{1991};\bibinfo{volume}{22}(\bibinfo{number}{2}):\bibinfo{pages}{425--434}.
%Type = Article
\bibitem[{Fick et~al.(2018)Fick, Petiet, Santin, Philippe, Lehericy, Deriche
  and Wassermann}]{fick2018non}
\bibinfo{author}{Fick\xfnm[ R.H.]}, \bibinfo{author}{Petiet\xfnm[ A.]},
  \bibinfo{author}{Santin\xfnm[ M.]}, \bibinfo{author}{Philippe\xfnm[ A.C.]},
  \bibinfo{author}{Lehericy\xfnm[ S.]}, \bibinfo{author}{Deriche\xfnm[ R.]},
  \bibinfo{author}{Wassermann\xfnm[ D.]}.
\newblock \bibinfo{title}{Non-parametric graphnet-regularized representation of
  dmri in space and time}.
\newblock \bibinfo{journal}{Medical image analysis}
  \bibinfo{year}{2018};\bibinfo{volume}{43}:\bibinfo{pages}{37--53}.
%Type = Inproceedings
\bibitem[{Fick et~al.(2015)Fick, Zucchelli, Girard, Descoteaux, Menegaz and
  Deriche}]{fick2015using}
\bibinfo{author}{Fick\xfnm[ R.H.J.]}, \bibinfo{author}{Zucchelli\xfnm[ M.]},
  \bibinfo{author}{Girard\xfnm[ G.]}, \bibinfo{author}{Descoteaux\xfnm[ M.]},
  \bibinfo{author}{Menegaz\xfnm[ G.]}, \bibinfo{author}{Deriche\xfnm[ R.]}.
\newblock \bibinfo{title}{Using 3d-shore and map-mri to obtain both
  tractography and microstructural constrast from a clinical dmri acquisition}.
\newblock In: \bibinfo{booktitle}{2015 IEEE 12th International Symposium on
  Biomedical Imaging (ISBI)}. \bibinfo{organization}{IEEE};
  \bibinfo{year}{2015}. p. \bibinfo{pages}{436--439}.
%Type = Article
\bibitem[{Filipiak et~al.(2022)Filipiak, Shepherd, Lin, Placantonakis, Boada
  and Baete}]{filipiak2022performance}
\bibinfo{author}{Filipiak\xfnm[ P.]}, \bibinfo{author}{Shepherd\xfnm[ T.]},
  \bibinfo{author}{Lin\xfnm[ Y.C.]}, \bibinfo{author}{Placantonakis\xfnm[
  D.G.]}, \bibinfo{author}{Boada\xfnm[ F.E.]}, \bibinfo{author}{Baete\xfnm[
  S.H.]}.
\newblock \bibinfo{title}{Performance of orientation distribution
  function-fingerprinting with a biophysical multicompartment diffusion model}.
\newblock \bibinfo{journal}{Magnetic Resonance in Medicine}
  \bibinfo{year}{2022};\bibinfo{volume}{88}(\bibinfo{number}{1}):\bibinfo{pages}{418--435}.
%Type = Article
\bibitem[{Fritz et~al.(2021)Fritz, Poser and Roebroeck}]{fritz2021mesmerised}
\bibinfo{author}{Fritz\xfnm[ F.]}, \bibinfo{author}{Poser\xfnm[ B.A.]},
  \bibinfo{author}{Roebroeck\xfnm[ A.]}.
\newblock \bibinfo{title}{{MESMERISED}: {S}uper-accelerating {T}1 relaxometry
  and diffusion {MRI} with {STEAM} at 7 {T} for quantitative multi-contrast and
  diffusion imaging}.
\newblock \bibinfo{journal}{Neuroimage}
  \bibinfo{year}{2021};\bibinfo{volume}{239}:\bibinfo{pages}{118285}.
%Type = Article
\bibitem[{Garyfallidis et~al.(2014)Garyfallidis, Brett, Amirbekian, Rokem, Van
  Der~Walt, Descoteaux, Nimmo-Smith and Contributors}]{garyfallidis2014dipy}
\bibinfo{author}{Garyfallidis\xfnm[ E.]}, \bibinfo{author}{Brett\xfnm[ M.]},
  \bibinfo{author}{Amirbekian\xfnm[ B.]}, \bibinfo{author}{Rokem\xfnm[ A.]},
  \bibinfo{author}{Van Der~Walt\xfnm[ S.]}, \bibinfo{author}{Descoteaux\xfnm[
  M.]}, \bibinfo{author}{Nimmo-Smith\xfnm[ I.]},
  \bibinfo{author}{Contributors\xfnm[ D.]}.
\newblock \bibinfo{title}{Dipy, a library for the analysis of diffusion mri
  data}.
\newblock \bibinfo{journal}{Frontiers in neuroinformatics}
  \bibinfo{year}{2014};\bibinfo{volume}{8}:\bibinfo{pages}{8}.
%Type = Article
\bibitem[{Golbabaee and Poon(2022)}]{golbabaee2022off}
\bibinfo{author}{Golbabaee\xfnm[ M.]}, \bibinfo{author}{Poon\xfnm[ C.]}.
\newblock \bibinfo{title}{An off-the-grid approach to multi-compartment
  magnetic resonance fingerprinting}.
\newblock \bibinfo{journal}{Inverse Problems}
  \bibinfo{year}{2022};\bibinfo{volume}{38}(\bibinfo{number}{8}):\bibinfo{pages}{085002}.
%Type = Inproceedings
\bibitem[{Grabner et~al.(2006)Grabner, Janke, Budge, Smith, Pruessner and
  Collins}]{grabner2006symmetric}
\bibinfo{author}{Grabner\xfnm[ G.]}, \bibinfo{author}{Janke\xfnm[ A.L.]},
  \bibinfo{author}{Budge\xfnm[ M.M.]}, \bibinfo{author}{Smith\xfnm[ D.]},
  \bibinfo{author}{Pruessner\xfnm[ J.]}, \bibinfo{author}{Collins\xfnm[ D.L.]}.
\newblock \bibinfo{title}{Symmetric atlasing and model based segmentation: an
  application to the hippocampus in older adults}.
\newblock In: \bibinfo{booktitle}{Medical Image Computing and Computer-Assisted
  Intervention--MICCAI 2006: 9th International Conference, Copenhagen, Denmark,
  October 1-6, 2006. Proceedings, Part II 9}. \bibinfo{organization}{Springer};
  \bibinfo{year}{2006}. p. \bibinfo{pages}{58--66}.
%Type = Article
\bibitem[{Haije et~al.(2020)Haije, {\"O}zarslan and
  Feragen}]{haije2020enforcing}
\bibinfo{author}{Haije\xfnm[ T.D.]}, \bibinfo{author}{{\"O}zarslan\xfnm[ E.]},
  \bibinfo{author}{Feragen\xfnm[ A.]}.
\newblock \bibinfo{title}{Enforcing necessary non-negativity constraints for
  common diffusion mri models using sum of squares programming}.
\newblock \bibinfo{journal}{NeuroImage}
  \bibinfo{year}{2020};\bibinfo{volume}{209}:\bibinfo{pages}{116405}.
%Type = Article
\bibitem[{Hosseinbor et~al.(2013)Hosseinbor, Chung, Wu and
  Alexander}]{hosseinbor2013bessel}
\bibinfo{author}{Hosseinbor\xfnm[ A.P.]}, \bibinfo{author}{Chung\xfnm[ M.K.]},
  \bibinfo{author}{Wu\xfnm[ Y.C.]}, \bibinfo{author}{Alexander\xfnm[ A.L.]}.
\newblock \bibinfo{title}{Bessel fourier orientation reconstruction (bfor): An
  analytical diffusion propagator reconstruction for hybrid diffusion imaging
  and computation of q-space indices}.
\newblock \bibinfo{journal}{NeuroImage}
  \bibinfo{year}{2013};\bibinfo{volume}{64}:\bibinfo{pages}{650--670}.
%Type = Article
\bibitem[{Hutter et~al.(2018)Hutter, Slator, Christiaens, Teixeira, Roberts,
  Jackson, Price, Malik and Hajnal}]{hutter2018integrated}
\bibinfo{author}{Hutter\xfnm[ J.]}, \bibinfo{author}{Slator\xfnm[ P.J.]},
  \bibinfo{author}{Christiaens\xfnm[ D.]}, \bibinfo{author}{Teixeira\xfnm[
  R.P.A.]}, \bibinfo{author}{Roberts\xfnm[ T.]}, \bibinfo{author}{Jackson\xfnm[
  L.]}, \bibinfo{author}{Price\xfnm[ A.N.]}, \bibinfo{author}{Malik\xfnm[ S.]},
  \bibinfo{author}{Hajnal\xfnm[ J.V.]}.
\newblock \bibinfo{title}{Integrated and efficient diffusion-relaxometry using
  {ZEBRA}}.
\newblock \bibinfo{journal}{Scientific reports}
  \bibinfo{year}{2018};\bibinfo{volume}{8}(\bibinfo{number}{1}):\bibinfo{pages}{15138}.
%Type = Article
\bibitem[{Jelescu et~al.(2020)Jelescu, Palombo, Bagnato and
  Schilling}]{jelescu2020challenges}
\bibinfo{author}{Jelescu\xfnm[ I.O.]}, \bibinfo{author}{Palombo\xfnm[ M.]},
  \bibinfo{author}{Bagnato\xfnm[ F.]}, \bibinfo{author}{Schilling\xfnm[ K.G.]}.
\newblock \bibinfo{title}{Challenges for biophysical modeling of
  microstructure}.
\newblock \bibinfo{journal}{Journal of Neuroscience Methods}
  \bibinfo{year}{2020};\bibinfo{volume}{344}:\bibinfo{pages}{108861}.
%Type = Article
\bibitem[{Jenkinson et~al.(2002)Jenkinson, Bannister, Brady and
  Smith}]{jenkinson2002improved}
\bibinfo{author}{Jenkinson\xfnm[ M.]}, \bibinfo{author}{Bannister\xfnm[ P.]},
  \bibinfo{author}{Brady\xfnm[ M.]}, \bibinfo{author}{Smith\xfnm[ S.]}.
\newblock \bibinfo{title}{Improved optimization for the robust and accurate
  linear registration and motion correction of brain images}.
\newblock \bibinfo{journal}{NeuroImage}
  \bibinfo{year}{2002};\bibinfo{volume}{17}(\bibinfo{number}{2}):\bibinfo{pages}{825--841}.
%Type = Article
\bibitem[{Jenkinson et~al.(2012)Jenkinson, Beckmann, Behrens, Woolrich and
  Smith}]{jenkinson2012fsl}
\bibinfo{author}{Jenkinson\xfnm[ M.]}, \bibinfo{author}{Beckmann\xfnm[ C.F.]},
  \bibinfo{author}{Behrens\xfnm[ T.E.]}, \bibinfo{author}{Woolrich\xfnm[
  M.W.]}, \bibinfo{author}{Smith\xfnm[ S.M.]}.
\newblock \bibinfo{title}{Fsl}.
\newblock \bibinfo{journal}{Neuroimage}
  \bibinfo{year}{2012};\bibinfo{volume}{62}(\bibinfo{number}{2}):\bibinfo{pages}{782--790}.
%Type = Article
\bibitem[{Jenkinson and Smith(2001)}]{jenkinson2001global}
\bibinfo{author}{Jenkinson\xfnm[ M.]}, \bibinfo{author}{Smith\xfnm[ S.]}.
\newblock \bibinfo{title}{A global optimisation method for robust affine
  registration of brain images}.
\newblock \bibinfo{journal}{Medical Image Analysis}
  \bibinfo{year}{2001};\bibinfo{volume}{5}(\bibinfo{number}{2}):\bibinfo{pages}{143--156}.
%Type = Article
\bibitem[{Jensen et~al.(2005)Jensen, Helpern, Ramani, Lu and
  Kaczynski}]{jensen2005diffusional}
\bibinfo{author}{Jensen\xfnm[ J.H.]}, \bibinfo{author}{Helpern\xfnm[ J.A.]},
  \bibinfo{author}{Ramani\xfnm[ A.]}, \bibinfo{author}{Lu\xfnm[ H.]},
  \bibinfo{author}{Kaczynski\xfnm[ K.]}.
\newblock \bibinfo{title}{Diffusional kurtosis imaging: the quantification of
  non-gaussian water diffusion by means of magnetic resonance imaging}.
\newblock \bibinfo{journal}{Magnetic Resonance in Medicine: An Official Journal
  of the International Society for Magnetic Resonance in Medicine}
  \bibinfo{year}{2005};\bibinfo{volume}{53}(\bibinfo{number}{6}):\bibinfo{pages}{1432--1440}.
%Type = Article
\bibitem[{Kim et~al.(2017)Kim, Doyle, Wisnowski, Kim and
  Haldar}]{kim2017diffusion}
\bibinfo{author}{Kim\xfnm[ D.]}, \bibinfo{author}{Doyle\xfnm[ E.K.]},
  \bibinfo{author}{Wisnowski\xfnm[ J.L.]}, \bibinfo{author}{Kim\xfnm[ J.H.]},
  \bibinfo{author}{Haldar\xfnm[ J.P.]}.
\newblock \bibinfo{title}{Diffusion-relaxation correlation spectroscopic
  imaging: a multidimensional approach for probing microstructure}.
\newblock \bibinfo{journal}{Magnetic resonance in medicine}
  \bibinfo{year}{2017};\bibinfo{volume}{78}(\bibinfo{number}{6}):\bibinfo{pages}{2236--2249}.
%Type = Article
\bibitem[{Lee et~al.(2021)Lee, Hyun, Lee, Choi, Shin, Min, Nam, Kim and
  Oh}]{lee2021so}
\bibinfo{author}{Lee\xfnm[ J.]}, \bibinfo{author}{Hyun\xfnm[ J.W.]},
  \bibinfo{author}{Lee\xfnm[ J.]}, \bibinfo{author}{Choi\xfnm[ E.J.]},
  \bibinfo{author}{Shin\xfnm[ H.G.]}, \bibinfo{author}{Min\xfnm[ K.]},
  \bibinfo{author}{Nam\xfnm[ Y.]}, \bibinfo{author}{Kim\xfnm[ H.J.]},
  \bibinfo{author}{Oh\xfnm[ S.H.]}.
\newblock \bibinfo{title}{So you want to image myelin using mri: an overview
  and practical guide for myelin water imaging}.
\newblock \bibinfo{journal}{Journal of Magnetic Resonance Imaging}
  \bibinfo{year}{2021};\bibinfo{volume}{53}(\bibinfo{number}{2}):\bibinfo{pages}{360--373}.
%Type = Article
\bibitem[{Leppert et~al.(2021)Leppert, Andrews, Campbell, Park, Pike, Polimeni
  and Tardif}]{leppert2021efficient}
\bibinfo{author}{Leppert\xfnm[ I.R.]}, \bibinfo{author}{Andrews\xfnm[ D.A.]},
  \bibinfo{author}{Campbell\xfnm[ J.S.]}, \bibinfo{author}{Park\xfnm[ D.J.]},
  \bibinfo{author}{Pike\xfnm[ G.B.]}, \bibinfo{author}{Polimeni\xfnm[ J.R.]},
  \bibinfo{author}{Tardif\xfnm[ C.L.]}.
\newblock \bibinfo{title}{Efficient whole-brain tract-specific {T}1 mapping at
  3{T} with slice-shuffled inversion-recovery diffusion-weighted imaging}.
\newblock \bibinfo{journal}{Magnetic Resonance in Medicine}
  \bibinfo{year}{2021};\bibinfo{volume}{86}(\bibinfo{number}{2}):\bibinfo{pages}{738--753}.
%Type = Article
\bibitem[{Merlet and Deriche(2013)}]{merlet2013continuous}
\bibinfo{author}{Merlet\xfnm[ S.L.]}, \bibinfo{author}{Deriche\xfnm[ R.]}.
\newblock \bibinfo{title}{Continuous diffusion signal, eap and odf estimation
  via compressive sensing in diffusion mri}.
\newblock \bibinfo{journal}{Medical image analysis}
  \bibinfo{year}{2013};\bibinfo{volume}{17}(\bibinfo{number}{5}):\bibinfo{pages}{556--572}.
%Type = Article
\bibitem[{Mori et~al.(2006)Mori, Wakana, Nagae-Poetscher and
  Van~Zijl}]{mori2006mri}
\bibinfo{author}{Mori\xfnm[ S.]}, \bibinfo{author}{Wakana\xfnm[ S.]},
  \bibinfo{author}{Nagae-Poetscher\xfnm[ L.]}, \bibinfo{author}{Van~Zijl\xfnm[
  P.]}.
\newblock \bibinfo{title}{{MRI} atlas of human white matter}.
\newblock \bibinfo{journal}{American Journal of Neuroradiology}
  \bibinfo{year}{2006};\bibinfo{volume}{27}(\bibinfo{number}{6}):\bibinfo{pages}{1384--1385}.
%Type = Article
\bibitem[{Nagtegaal et~al.(2020)Nagtegaal, Koken, Amthor and
  Doneva}]{nagtegaal2020fast}
\bibinfo{author}{Nagtegaal\xfnm[ M.]}, \bibinfo{author}{Koken\xfnm[ P.]},
  \bibinfo{author}{Amthor\xfnm[ T.]}, \bibinfo{author}{Doneva\xfnm[ M.]}.
\newblock \bibinfo{title}{Fast multi-component analysis using a joint sparsity
  constraint for mr fingerprinting}.
\newblock \bibinfo{journal}{Magnetic resonance in medicine}
  \bibinfo{year}{2020};\bibinfo{volume}{83}(\bibinfo{number}{2}):\bibinfo{pages}{521--534}.
%Type = Article
\bibitem[{Ning et~al.(2019)Ning, Gagoski, Szczepankiewicz, Westin and
  Rathi}]{ning2019joint}
\bibinfo{author}{Ning\xfnm[ L.]}, \bibinfo{author}{Gagoski\xfnm[ B.]},
  \bibinfo{author}{Szczepankiewicz\xfnm[ F.]}, \bibinfo{author}{Westin\xfnm[
  C.F.]}, \bibinfo{author}{Rathi\xfnm[ Y.]}.
\newblock \bibinfo{title}{Joint relaxation-diffusion imaging moments to probe
  neurite microstructure}.
\newblock \bibinfo{journal}{IEEE transactions on medical imaging}
  \bibinfo{year}{2019};\bibinfo{volume}{39}(\bibinfo{number}{3}):\bibinfo{pages}{668--677}.
%Type = Article
\bibitem[{Ning et~al.(2015)Ning, Westin and Rathi}]{ning2015estimating}
\bibinfo{author}{Ning\xfnm[ L.]}, \bibinfo{author}{Westin\xfnm[ C.F.]},
  \bibinfo{author}{Rathi\xfnm[ Y.]}.
\newblock \bibinfo{title}{Estimating diffusion propagator and its moments using
  directional radial basis functions}.
\newblock \bibinfo{journal}{IEEE transactions on medical imaging}
  \bibinfo{year}{2015};\bibinfo{volume}{34}(\bibinfo{number}{10}):\bibinfo{pages}{2058--2078}.
%Type = Inproceedings
\bibitem[{{\"O}zarslan et~al.(2009){\"O}zarslan, Koay, Shepherd, Blackband and
  Basser}]{ozarslan2009simple}
\bibinfo{author}{{\"O}zarslan\xfnm[ E.]}, \bibinfo{author}{Koay\xfnm[ C.]},
  \bibinfo{author}{Shepherd\xfnm[ T.M.]}, \bibinfo{author}{Blackband\xfnm[
  S.J.]}, \bibinfo{author}{Basser\xfnm[ P.J.]}.
\newblock \bibinfo{title}{Simple harmonic oscillator based reconstruction and
  estimation for three-dimensional q-space mri}.
\newblock In: \bibinfo{booktitle}{Proc. Intl. Soc. Mag. Reson. Med}.
  \bibinfo{organization}{Citeseer}; volume~\bibinfo{volume}{17};
  \bibinfo{year}{2009}. p. \bibinfo{pages}{1396}.
%Type = Article
\bibitem[{{\"O}zarslan et~al.(2013){\"O}zarslan, Koay, Shepherd, Komlosh,
  {\.I}rfano{\u{g}}lu, Pierpaoli and Basser}]{ozarslan2013mean}
\bibinfo{author}{{\"O}zarslan\xfnm[ E.]}, \bibinfo{author}{Koay\xfnm[ C.G.]},
  \bibinfo{author}{Shepherd\xfnm[ T.M.]}, \bibinfo{author}{Komlosh\xfnm[
  M.E.]}, \bibinfo{author}{{\.I}rfano{\u{g}}lu\xfnm[ M.O.]},
  \bibinfo{author}{Pierpaoli\xfnm[ C.]}, \bibinfo{author}{Basser\xfnm[ P.J.]}.
\newblock \bibinfo{title}{Mean apparent propagator (map) mri: a novel diffusion
  imaging method for mapping tissue microstructure}.
\newblock \bibinfo{journal}{NeuroImage}
  \bibinfo{year}{2013};\bibinfo{volume}{78}:\bibinfo{pages}{16--32}.
%Type = Article
\bibitem[{Pieciak et~al.(2023)Pieciak, Par{\'\i}s, Beck, Maximov,
  Trist{\'a}n-Vega, de~Luis-Garc{\'\i}a, Westlye and
  Aja-Fern{\'a}ndez}]{pieciak2023spherical}
\bibinfo{author}{Pieciak\xfnm[ T.]}, \bibinfo{author}{Par{\'\i}s\xfnm[ G.]},
  \bibinfo{author}{Beck\xfnm[ D.]}, \bibinfo{author}{Maximov\xfnm[ I.I.]},
  \bibinfo{author}{Trist{\'a}n-Vega\xfnm[ A.]},
  \bibinfo{author}{de~Luis-Garc{\'\i}a\xfnm[ R.]},
  \bibinfo{author}{Westlye\xfnm[ L.T.]},
  \bibinfo{author}{Aja-Fern{\'a}ndez\xfnm[ S.]}.
\newblock \bibinfo{title}{Spherical means-based free-water volume fraction from
  diffusion {MRI} increases non-linearly with age in the white matter of the
  healthy human brain}.
\newblock \bibinfo{journal}{NeuroImage}
  \bibinfo{year}{2023};\bibinfo{volume}{279}:\bibinfo{pages}{120324}.
%Type = Inproceedings
\bibitem[{Pizzolato et~al.(2020)Pizzolato, Palombo, Bonet-Carne, Tax, Grussu,
  Ianus, Bogusz, Pieciak, Ning, Larochelle et~al.}]{pizzolato2020acquiring}
\bibinfo{author}{Pizzolato\xfnm[ M.]}, \bibinfo{author}{Palombo\xfnm[ M.]},
  \bibinfo{author}{Bonet-Carne\xfnm[ E.]}, \bibinfo{author}{Tax\xfnm[ C.M.]},
  \bibinfo{author}{Grussu\xfnm[ F.]}, \bibinfo{author}{Ianus\xfnm[ A.]},
  \bibinfo{author}{Bogusz\xfnm[ F.]}, \bibinfo{author}{Pieciak\xfnm[ T.]},
  \bibinfo{author}{Ning\xfnm[ L.]}, \bibinfo{author}{Larochelle\xfnm[ H.]},
  et~al.
\newblock \bibinfo{title}{Acquiring and predicting multidimensional diffusion
  (mudi) data: an open challenge}.
\newblock In: \bibinfo{booktitle}{Computational Diffusion MRI: MICCAI Workshop,
  Shenzhen, China, October 2019}. \bibinfo{organization}{Springer};
  \bibinfo{year}{2020}. p. \bibinfo{pages}{195--208}.
%Type = Article
\bibitem[{Slator et~al.(2021{\natexlab{a}})Slator, Hutter, Marinescu, Palombo,
  Jackson, Ho, Chappell, Rutherford, Hajnal and Alexander}]{slator2021data}
\bibinfo{author}{Slator\xfnm[ P.J.]}, \bibinfo{author}{Hutter\xfnm[ J.]},
  \bibinfo{author}{Marinescu\xfnm[ R.V.]}, \bibinfo{author}{Palombo\xfnm[ M.]},
  \bibinfo{author}{Jackson\xfnm[ L.H.]}, \bibinfo{author}{Ho\xfnm[ A.]},
  \bibinfo{author}{Chappell\xfnm[ L.C.]}, \bibinfo{author}{Rutherford\xfnm[
  M.]}, \bibinfo{author}{Hajnal\xfnm[ J.V.]}, \bibinfo{author}{Alexander\xfnm[
  D.C.]}.
\newblock \bibinfo{title}{Data-driven multi-contrast spectral microstructure
  imaging with inspect: Integrated spectral component estimation and mapping}.
\newblock \bibinfo{journal}{Medical image analysis}
  \bibinfo{year}{2021}{\natexlab{a}};\bibinfo{volume}{71}:\bibinfo{pages}{102045}.
%Type = Article
\bibitem[{Slator et~al.(2021{\natexlab{b}})Slator, Palombo, Miller, Westin,
  Laun, Kim, Haldar, Benjamini, Lemberskiy, de~Almeida~Martins
  et~al.}]{slator2021combined}
\bibinfo{author}{Slator\xfnm[ P.J.]}, \bibinfo{author}{Palombo\xfnm[ M.]},
  \bibinfo{author}{Miller\xfnm[ K.L.]}, \bibinfo{author}{Westin\xfnm[ C.F.]},
  \bibinfo{author}{Laun\xfnm[ F.]}, \bibinfo{author}{Kim\xfnm[ D.]},
  \bibinfo{author}{Haldar\xfnm[ J.P.]}, \bibinfo{author}{Benjamini\xfnm[ D.]},
  \bibinfo{author}{Lemberskiy\xfnm[ G.]},
  \bibinfo{author}{de~Almeida~Martins\xfnm[ J.P.]}, et~al.
\newblock \bibinfo{title}{Combined diffusion-relaxometry microstructure
  imaging: Current status and future prospects}.
\newblock \bibinfo{journal}{Magnetic resonance in medicine}
  \bibinfo{year}{2021}{\natexlab{b}};\bibinfo{volume}{86}(\bibinfo{number}{6}):\bibinfo{pages}{2987--3011}.
%Type = Article
\bibitem[{Tang et~al.(2018)Tang, Fernandez-Granda, Lannuzel, Bernstein,
  Lattanzi, Cloos, Knoll and Assl{\"a}nder}]{tang2018multicompartment}
\bibinfo{author}{Tang\xfnm[ S.]}, \bibinfo{author}{Fernandez-Granda\xfnm[ C.]},
  \bibinfo{author}{Lannuzel\xfnm[ S.]}, \bibinfo{author}{Bernstein\xfnm[ B.]},
  \bibinfo{author}{Lattanzi\xfnm[ R.]}, \bibinfo{author}{Cloos\xfnm[ M.]},
  \bibinfo{author}{Knoll\xfnm[ F.]}, \bibinfo{author}{Assl{\"a}nder\xfnm[ J.]}.
\newblock \bibinfo{title}{Multicompartment magnetic resonance fingerprinting}.
\newblock \bibinfo{journal}{Inverse problems}
  \bibinfo{year}{2018};\bibinfo{volume}{34}(\bibinfo{number}{9}):\bibinfo{pages}{094005}.
%Type = Article
\bibitem[{Tibshirani et~al.(2005)Tibshirani, Saunders, Rosset, Zhu and
  Knight}]{tibshirani2005sparsity}
\bibinfo{author}{Tibshirani\xfnm[ R.]}, \bibinfo{author}{Saunders\xfnm[ M.]},
  \bibinfo{author}{Rosset\xfnm[ S.]}, \bibinfo{author}{Zhu\xfnm[ J.]},
  \bibinfo{author}{Knight\xfnm[ K.]}.
\newblock \bibinfo{title}{Sparsity and smoothness via the fused lasso}.
\newblock \bibinfo{journal}{Journal of the Royal Statistical Society Series B:
  Statistical Methodology}
  \bibinfo{year}{2005};\bibinfo{volume}{67}(\bibinfo{number}{1}):\bibinfo{pages}{91--108}.
%Type = Article
\bibitem[{Tobisch et~al.(2019)Tobisch, Schultz, Stirnberg, Varela-Mattatall,
  Knutsson, Irarr{\'a}zaval and St{\"o}cker}]{tobisch2019comparison}
\bibinfo{author}{Tobisch\xfnm[ A.]}, \bibinfo{author}{Schultz\xfnm[ T.]},
  \bibinfo{author}{Stirnberg\xfnm[ R.]},
  \bibinfo{author}{Varela-Mattatall\xfnm[ G.]}, \bibinfo{author}{Knutsson\xfnm[
  H.]}, \bibinfo{author}{Irarr{\'a}zaval\xfnm[ P.]},
  \bibinfo{author}{St{\"o}cker\xfnm[ T.]}.
\newblock \bibinfo{title}{Comparison of basis functions and q-space sampling
  schemes for robust compressed sensing reconstruction accelerating diffusion
  spectrum imaging}.
\newblock \bibinfo{journal}{NMR in Biomedicine}
  \bibinfo{year}{2019};\bibinfo{volume}{32}(\bibinfo{number}{3}):\bibinfo{pages}{e4055}.
%Type = Article
\bibitem[{Trist{\'a}n-Vega et~al.(2023)Trist{\'a}n-Vega, Pieciak, Par{\'\i}s,
  Rodr{\'\i}guez-Galv{\'a}n and Aja-Fern{\'a}ndez}]{tristan2023hydi}
\bibinfo{author}{Trist{\'a}n-Vega\xfnm[ A.]}, \bibinfo{author}{Pieciak\xfnm[
  T.]}, \bibinfo{author}{Par{\'\i}s\xfnm[ G.]},
  \bibinfo{author}{Rodr{\'\i}guez-Galv{\'a}n\xfnm[ J.R.]},
  \bibinfo{author}{Aja-Fern{\'a}ndez\xfnm[ S.]}.
\newblock \bibinfo{title}{{HYDI}-{DSI} revisited: {C}onstrained non-parametric
  {EAP} imaging without q-space re-gridding}.
\newblock \bibinfo{journal}{Medical Image Analysis}
  \bibinfo{year}{2023};\bibinfo{volume}{84}:\bibinfo{pages}{102728}.
%Type = Article
\bibitem[{Tuch(2004)}]{tuch2004q}
\bibinfo{author}{Tuch\xfnm[ D.S.]}.
\newblock \bibinfo{title}{Q-ball imaging}.
\newblock \bibinfo{journal}{Magnetic Resonance in Medicine}
  \bibinfo{year}{2004};\bibinfo{volume}{52}(\bibinfo{number}{6}):\bibinfo{pages}{1358--1372}.
%Type = Article
\bibitem[{Viola and Wells~III(1997)}]{viola1997alignment}
\bibinfo{author}{Viola\xfnm[ P.]}, \bibinfo{author}{Wells~III\xfnm[ W.M.]}.
\newblock \bibinfo{title}{Alignment by maximization of mutual information}.
\newblock \bibinfo{journal}{International journal of computer vision}
  \bibinfo{year}{1997};\bibinfo{volume}{24}(\bibinfo{number}{2}):\bibinfo{pages}{137--154}.
%Type = Article
\bibitem[{Vos et~al.(2011)Vos, Jones, Viergever and Leemans}]{vos2011partial}
\bibinfo{author}{Vos\xfnm[ S.B.]}, \bibinfo{author}{Jones\xfnm[ D.K.]},
  \bibinfo{author}{Viergever\xfnm[ M.A.]}, \bibinfo{author}{Leemans\xfnm[ A.]}.
\newblock \bibinfo{title}{Partial volume effect as a hidden covariate in dti
  analyses}.
\newblock \bibinfo{journal}{Neuroimage}
  \bibinfo{year}{2011};\bibinfo{volume}{55}(\bibinfo{number}{4}):\bibinfo{pages}{1566--1576}.
%Type = Article
\bibitem[{Wang et~al.(2024)Wang, Zhu, Luo and He}]{wang2024q}
\bibinfo{author}{Wang\xfnm[ Y.]}, \bibinfo{author}{Zhu\xfnm[ Y.]},
  \bibinfo{author}{Luo\xfnm[ L.]}, \bibinfo{author}{He\xfnm[ J.]}.
\newblock \bibinfo{title}{Q-space imaging based on {G}aussian radial basis
  function with {L}aplace regularization}.
\newblock \bibinfo{journal}{Magnetic Resonance in Medicine}
  \bibinfo{year}{2024};\bibinfo{volume}{92}(\bibinfo{number}{1}):\bibinfo{pages}{128--144}.
%Type = Article
\bibitem[{Zucchelli et~al.(2016)Zucchelli, Brusini, M{\'e}ndez, Daducci,
  Granziera and Menegaz}]{zucchelli2016lies}
\bibinfo{author}{Zucchelli\xfnm[ M.]}, \bibinfo{author}{Brusini\xfnm[ L.]},
  \bibinfo{author}{M{\'e}ndez\xfnm[ C.A.]}, \bibinfo{author}{Daducci\xfnm[
  A.]}, \bibinfo{author}{Granziera\xfnm[ C.]}, \bibinfo{author}{Menegaz\xfnm[
  G.]}.
\newblock \bibinfo{title}{What lies beneath? {D}iffusion {EAP}-based study of
  brain tissue microstructure}.
\newblock \bibinfo{journal}{Medical image analysis}
  \bibinfo{year}{2016};\bibinfo{volume}{32}:\bibinfo{pages}{145--156}.

\end{thebibliography}

\end{document}